\documentclass[superscriptaddress,showpacs,amssymb,10pt,reprint,aps,prd,longbibliography,nofootinbib,floatfix]{revtex4-2}

\usepackage{graphicx,epsfig,amssymb,times} 
\usepackage{amsmath,amsfonts}
\usepackage{bm}
\usepackage{epstopdf}
\usepackage{hyperref}
\usepackage[caption=false]{subfig}
\usepackage[usenames]{color}   
\usepackage[dvipsnames]{xcolor}
\usepackage[titletoc]{appendix}

\usepackage[normalem]{ulem}

\definecolor{coolblack}{rgb}{0.0, 0.18, 0.39}
\definecolor{darkred}{rgb}{0.5,0,0}
\definecolor{darkgreen}{rgb}{0,0.5,0}
\definecolor{darkblue}{rgb}{0,0,0.5}
\definecolor{lapislazuli}{rgb}{0.15, 0.38, 0.61}
\definecolor{venetianred}{rgb}{0.78, 0.03, 0.08}
\definecolor{bleudefrance}{rgb}{0.19, 0.55, 0.91}
\definecolor{dogwoodrose}{rgb}{0.84, 0.09, 0.41}
\hypersetup{colorlinks=true, citecolor=darkblue, linkcolor=darkblue, 
urlcolor = darkblue}
\begin{document}
\title{Scattering properties of charged black holes in nonlinear and Maxwell's electrodynamics}
	
	\author{Marco A. A. de Paula}
	\email{marco.paula@icen.ufpa.br}
	\affiliation{Programa de P\'os-Gradua\c{c}\~{a}o em F\'{\i}sica, Universidade 
		Federal do Par\'a, 66075-110, Bel\'em, Par\'a, Brazil.}

\author{Luiz C. S. Leite}
\email{luiz.leite@ifpa.edu.br}
\affiliation{Campus Altamira, Instituto Federal do Par\'a, 68377-630, Altamira, Par\'a, Brazil.}
	
\author{Lu\'is C. B. Crispino}
\email{crispino@ufpa.br}
\affiliation{Programa de P\'os-Gradua\c{c}\~{a}o em F\'{\i}sica, Universidade 
		Federal do Par\'a, 66075-110, Bel\'em, Par\'a, Brazil.}

\begin{abstract}
We investigate the scattering properties of a massless scalar field in the background of a charged Ay\'on-Beato-Garc\'ia regular black hole solution. Using a numerical approach, we compute the differential scattering cross section for arbitrary values of the scattering angle and of the incident wave frequency. We compare our results with those obtained via the classical geodesic scattering of massless particles, as well as with the semiclassical glory approximation, and show that they present an excellent agreement in the corresponding limits. We also show that Ay\'on-Beato-Garc\'ia and Reissner-Nordstr\"om black hole solutions present similar scattering properties, for low-to-moderate values of the black hole electric charge, for any value of the scattering angle.
\end{abstract}

\date{\today}

\maketitle

\section{Introduction}
Scattering is one of the most common procedures in Physics to improve our knowledge about objects. In black hole (BH) scenarios, this procedure is a little more sophisticated due to the existence of a one-way membrane, called event horizon. Thus, for instance, we cannot investigate the interior of BHs, as we study the interior of the atomic nucleus. Nevertheless, it is possible to analyze how BHs absorb and scatter bosonic and fermionic fields, propagating in their vicinities. Typically, in order to do that, we choose a particular field with spin $s$, e.g., scalar ($s = 0$), Dirac ($s = 1/2$), electromagnetic ($s = 1$) or gravitational ($s = 2$) field, and study its absorption and scattering processes in a given BH spacetime. Since the late 1960s this has been done in several standard BHs scenarios (cf., e.g., Refs.~\cite{FHM1988,CDE2009,OCH2011,CDHO2014,CDHO2015,BC2016,LBC2017} and references therein).

Once there are several observational evidences that BHs are likely to be surrounded by matter~\cite{N2005}, the investigation of how they interact with their astrophysical environment becomes essential. Moreover, the scattering of light plays an important role in the analysis of BHs shadows. As an example, we may anticipate subtle properties of BHs shadows which, in principle, could be measured by the Event Horizon Telescope Consortium (EHTC), which recently unveiled the first shadow image of a supermassive BH~\cite{AAA2019L1}. We can, e.g., compute the shadows of distinct BH solutions and verify if they can be distinguishable through the EHTC data (cf., e.g., Ref.~\cite{SVX2020}).

In the last years, the absorption and scattering properties of regular BHs (RBHs) solutions (for a review about RBHs see, e.g., Ref.~\cite{A2008}) have gained attention~\cite{MC2014,MOC2015,S2017,SBP2018,PLC2020}. These solutions are singularity-free BHs, initially proposed as toy models within GR, as it has been the case of the first RBH line element proposed by James Bardeen in 1968~\cite{B1968}. The first charged RBH solutions associated with a source were obtained only in the late 1990s, based on the minimal coupling between gravity and nonlinear electrodynamics (NED) models~\cite{ABG1998,ABG1999,ABG1999-2,ABG2000}. Presently, several RBH solutions based on a NED framework have already been proposed (cf., e.g., Refs.~\cite{B2001,D2004,M2004,BV2014,M2015,K2017,JRH2015,SR2018} and references therein). 

One of the first NED theories was proposed by Max Born and Leopold Infeld  in 1934~\cite{B1934,BI1934} as a possible generalization of Maxwell's theory for strong electromagnetic fields. Their key motivation was to find a finite self-energy for the electromagnetic field of a point charge, in contrast with the infinite result obtained via Maxwell's theory. The NED theory has several applications in physics, beyond helping to obtain new BH solutions, especially in cosmology~\cite{VAL2002,NBS2004,RPM2016}, quantum electrodynamics~\cite{D2003,ATLAS2017,ATLAS2019,PVLAS2020,A2021}, and string/M theories~\cite{FT1985,SW1999,A2000}. In the so-called $P$ framework~\cite{B2001}, the action that governs the dynamics of NED theory minimally coupled to GR is given by
\begin{equation}
\label{S}\mathrm{S} = \dfrac{1}{4 \pi}\int d^{4}x \sqrt{-g} \left[ \dfrac{1}{4}R-\left(2P\mathcal{H}_{P}-\mathcal{H}(P)\right) \right], 
\end{equation}
where $g$ is the determinant of the metric tensor $g_{\mu \nu}$, $R$ is the corresponding Ricci scalar, $\mathcal{H}(P)$ is a structural function, and $\mathcal{H}_{P} \equiv \partial\mathcal{H}/\partial P$. The function $P$ is defined as $P \equiv (1/4)P_{\mu \nu}P^{\mu \nu}$, where $P_{\mu \nu}$ is an anti-symmetric tensor. The $P$ framework is useful to derive exact solutions of the Einstein-field equations with nonlinear electromagnetic sources~\cite{HGP1987}, especially electrically charged NED-based RBHs solutions~\cite{BV2014}.

We investigate the scattering cross section (SCS) of massless test scalar fields in the background of an electrically charged Ay\'on-Beato-Garc\'ia (ABG) RBH spacetime~\cite{ABG1998}. This solution was obtained by solving the corresponding field equations associated with the action~\eqref{S} for
\begin{equation}
\label{SOURCE}\mathcal{H}(P) = P\dfrac{\left(1-3\sqrt{-2Q^{2}P}\right)}{(1+\sqrt{-2Q^{2}P})^{3}}-\dfrac{3M}{|Q|^{3}}\left(\dfrac{\sqrt{-2Q^{2}P}}{1+\sqrt{-2Q^{2}P}}\right)^\frac{5}{2},
\end{equation}
where $M$ and $Q$ represent the mass and the electric charge of the central object, respectively, with $P = -(Q^{2}/2r^{4})$. The line element of the ABG solution is static and spherically symmetric, and can be written as
\begin{equation}
\label{LE} ds^{2}= f(r)dt^{2}-f(r)^{-1}dr^{2}-r^{2}\left( d\theta^2 + \sin^2\theta\, d\varphi^2\right),
\end{equation}
with the metric function $f(r)$ given by
\begin{equation}
\label{MF_ABG}f(r) = f^{\rm{ABG}}(r) \equiv 1-\frac{2Mr^{2}}{(r^{2}+Q^{2})^{3/2}}+\frac{Q^{2}r^{2}}{(r^{2}+Q^{2})^{2}}. 
\end{equation}
Here we shall call the spacetime configuration~\eqref{LE}, \eqref{MF_ABG} as the ABG solution.~\footnote{It is worth noting that Ay\'on-Beato and Garc\'ia also derived other two similar RBHs solutions~\cite{ABG1999,ABG1999-2}.} Moreover, the ABG solution behaves asymptotically as the Reissner-Nordstr\"om (RN) solution, i.e.,
\begin{equation}
\label{MF_RN}\lim_{r \rightarrow \infty} f^{\rm{ABG}}(r) \rightarrow f^{\rm{RN}}(r) \equiv 1 - \dfrac{2M}{r} + \dfrac{Q^{2}}{r^{2}},
\end{equation}
with $f^{\rm{RN}}(r)$ being the metric function of the RN spacetime. We also address the possibility of RBHs solutions mimicking the well-known scattering properties of standard BHs.

The ABG solution describes RBHs when $|Q| \leq Q_{\rm{ext}} \approx 0.6341M$. For $|Q| < Q_{\rm{ext}}$ the solution presents two horizons, while for $|Q| = Q_{\rm{ext}}$ we have the so-called extreme ABG RBH, for which the two horizons degenerate into a single one. On the other hand, $|Q| > Q_{\rm{ext}}$ is associated to horizonless solutions. Here we shall restrict our analysis to BHs, for which the condition $|Q| \leq Q_{\rm{ext}}$ holds, and we exhibit our results in terms of the normalized electric charge $\alpha \equiv Q/Q_{\rm{ext}}$, which satisfies $0 \leq \alpha \leq 1$. A detailed analysis of the main differences between ABG and RN spacetimes was performed in Ref.~\cite{PLC2020}. 

The remainder of this paper is organized as follows. In Sect.~\ref{sec:css} we study the classical and semiclassical scattering. The propagation of a massless test scalar field in the background of the ABG solution, as well as the SCS, are considered in Sect.~\ref{sec:ws}. We present a selection of our numerical results concerning the scalar field scattering by ABG RBHs in Sect.~\ref{sec:rst}, and our final remarks are presented in Sect.~\ref{sec:fr}. Throughout this paper we use the natural units, for which $G = c = \hbar = 1$, and the metric signature ($+,-,-,-$).

\section{Classical and Semiclassical Scattering}\label{sec:css}
The partial-waves method allows the computation of the SCS for arbitrary values of the scattered wave frequency and of the scattering angle. However, it is possible to anticipate some features of the SCS, considering high-frequency incident waves, for small and large scattering angles, using classical and semiclassical analyses, respectively. In this section, we investigate the scattering of null geodesics\footnote{The classical analysis of null geodesics presented in this section, regarding the ABG RBH, is not concerned to photons, but rather to massless particles with nature other than electromagnetic, since in NED theories photons propagate along null geodesics of an effective geometry~\cite{GDP1981,MN2000}.}, as well as the semiclassical glory scattering, by ABG RBHs.

\subsection{Geodesic Scattering}\label{subsec:gs}
\begin{figure}[!htbp]
\begin{centering}
    \includegraphics[width=\columnwidth]{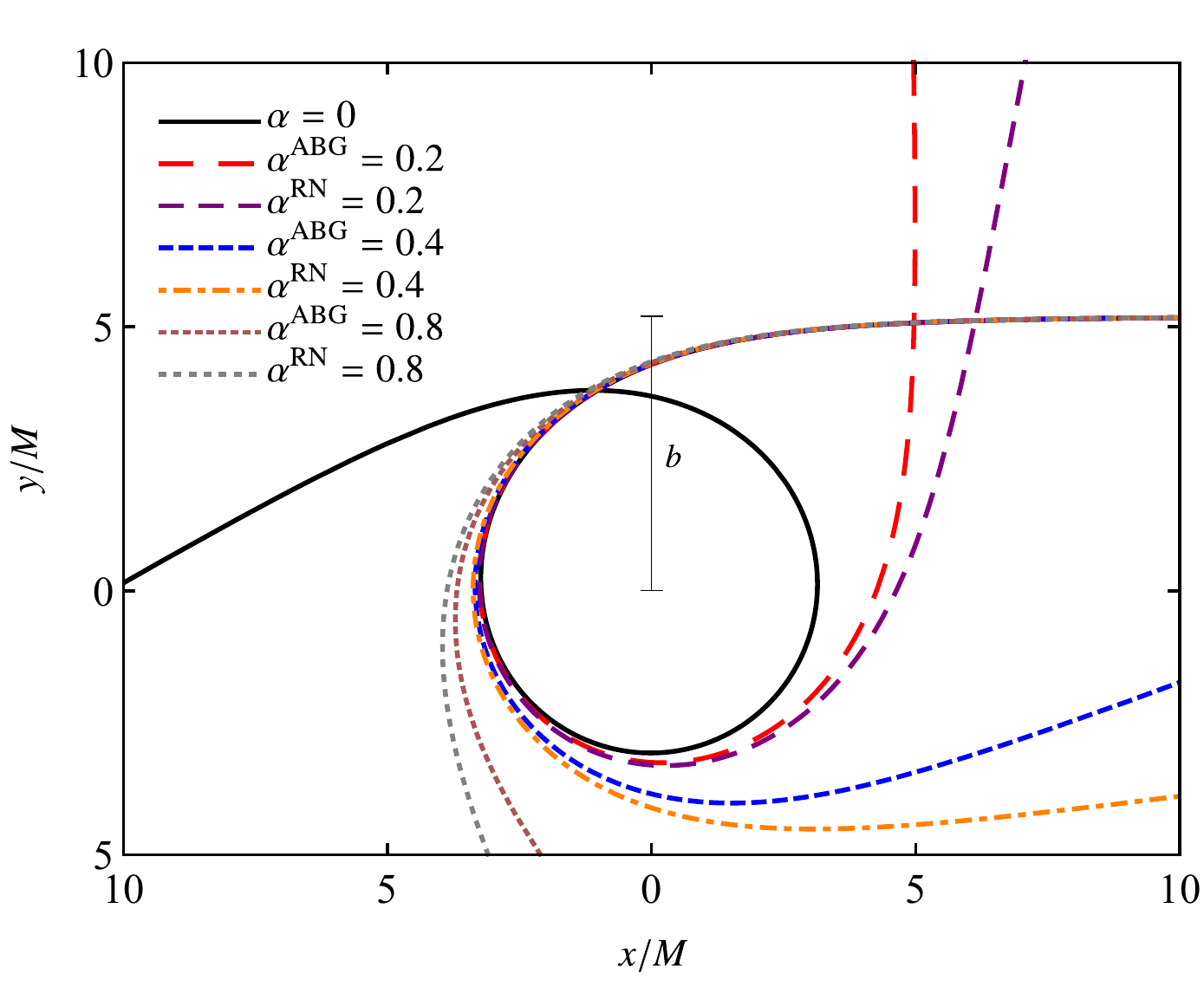}
    \caption{Parametric plot of geodesics approaching ABG and RN BHs from infinity, considering different values of $\alpha$. Here we set $b = 5.2M$, and we choose the numerical infinity at $r_{\infty} = 100M$. The Schwarzschild case ($\alpha = 0$) is also exhibited, for comparison}
    \label{gs}
\end{centering}
\end{figure}

The classical Lagrangian, $\mathrm{L_{\rm{geo}}}$, associated with the geodesic motion can be written as
\begin{equation}
\label{L}\mathrm{L_{\rm{geo}}} = \dfrac{1}{2} g_{\mu\nu}\dot{x}^{\mu}\dot{x}^{\nu},
\end{equation}
where the overdot denotes differentiation with respect to an affine parameter. Since the line element~\eqref{LE} does not depend explicitly on the coordinates $t$ and $\phi$, we obtain two constants of motion 
\begin{equation}
\label{CQ}E = f(r)\dot{t} \ \ \ \text{and} \ \ \ L = r^{2}\dot{\varphi},
\end{equation}
corresponding to the energy and the angular momentum of the particle, respectively; where we consider, without loss of generality, the motion in the equatorial plane, i.e., $\theta = \pi/2$.
For null geodesics, the condition $\mathrm{L_{\rm{geo}}} = 0$ is satisfied, and we can show that
\begin{equation}
\label{RME}\left(\dfrac{du}{d\varphi}\right)^{2} = \mathcal{T}(u) \equiv \dfrac{1}{b^{2}}-f(1/u) u^{2},
\end{equation}
in which $u \equiv 1/r$, and $b \equiv L/E$ is the impact parameter. By solving Eq.~\eqref{RME} with suitable boundary conditions, we can obtain the geodesics associated to massless particles. In Fig.~\ref{gs} we exhibit some of these geodesics for ABG and RN spacetimes. Wee see that the smaller the value of $\alpha$ is, the stronger is the influence of the BH in the massless particle trajectory. 

We can define the critical radius, $r_{c} = 1/u_{c}$, i.e., the radius of the unstable null circular orbit, and the corresponding critical impact parameter $b_{c}$. It follows that $\mathcal{T}(u)|_{u = u_{c}} = 0$ and $\frac{d\mathcal{T}(u)}{du}\big|_{u = u_{c}} = 0$, and we get the following pair of equations
\begin{align}
\label{CIP}b_{c} = \dfrac{L_{c}}{E_{c}} &= \dfrac{1}{u_{c}\sqrt{f(1/u_{c})}},\\
\label{CR}2f(1/u_{c}) & +  u_{c}\dfrac{f(1/u_{c})}{du} = 0.
\end{align}
Massless particles coming from infinity with impact parameter $b = b_c $ orbit around the BH an infinite number of times at $r = r_c$. When $b < b_{c}$ they will cross the event horizon, being absorbed by the BH. On the other hand, for $b > b_{c}$ massless particles will be scattered by the central object, with a turning point $u_0$ that satisfies $\mathcal{T}(u)|_{u = u_{0}} = 0$, corresponding to the radius of maximum approximation of the particle, for a given value of $b$. In the latter case, the deflection angle of the scattered massless particle is given by
\begin{equation}
\label{DA}\Theta(b) = 2\gamma(b) - \pi,
\end{equation} 
where
\begin{equation}
\gamma(b) = \int_{0}^{u_{0}}\dfrac{1}{\sqrt{\mathcal{T}(u)}}du.
\end{equation}

The classical differential SCS is given by~\cite{N2013}
\begin{equation}
\label{CSCS}\dfrac{d\sigma_{\rm{cl}}}{d\Omega} = \dfrac{b}{\sin\theta}\bigg|\dfrac{db}{d\Theta}\bigg|,
\end{equation}
where $\theta$ is the scattering angle, which is related to the deflection angle by $\Theta=\theta+2n\pi$ with $n\in\mathbb{Z}^{+}$. Note that Eq.~\eqref{CSCS} takes into account the number of times, $n$, that a massless particle orbits the BH before being scattered to infinity (cf., e. g., the black solid line in Fig.~\ref{gs}, for which $n = 1$). The classical SCS may be obtained by inverting Eq.~\eqref{DA}, and inserting $b(\Theta)$ into Eq.~\eqref{CSCS}.

Once the ABG RBH solution tends to the RN geometry in the weak-field limit, it is expected that, in this limit, the deflection angle, in the first order, reduces to~\cite{CDE2009}
\begin{equation}
\label{DA_WFL}\Theta(b) \approx \dfrac{4M}{b},
\end{equation}
which is Einstein's deflection angle~\cite{W1984}. 
This can be confirmed using the Gauss-Bonnet theorem~\cite{GW2008,JHO2020,FZL2021}. Hence, considering the weak-field expansion of the deflection angle, the classical differential SCS for small scattering angles reads
\begin{equation}
\label{CSCS_SA}\dfrac{d\sigma_{\rm{cl}}}{d\Omega} \approx \dfrac{16M^{2}}{\theta^{4}}.
\end{equation}
From Eqs.~\eqref{DA_WFL} and~\eqref{CSCS_SA} we see that, for small values of $\theta$, the dominant term of the deflection angle and of the classical differential SCS is not modified by the presence of the BH charge.

\begin{figure}[!htbp]
\begin{centering}
    \includegraphics[width=\columnwidth]{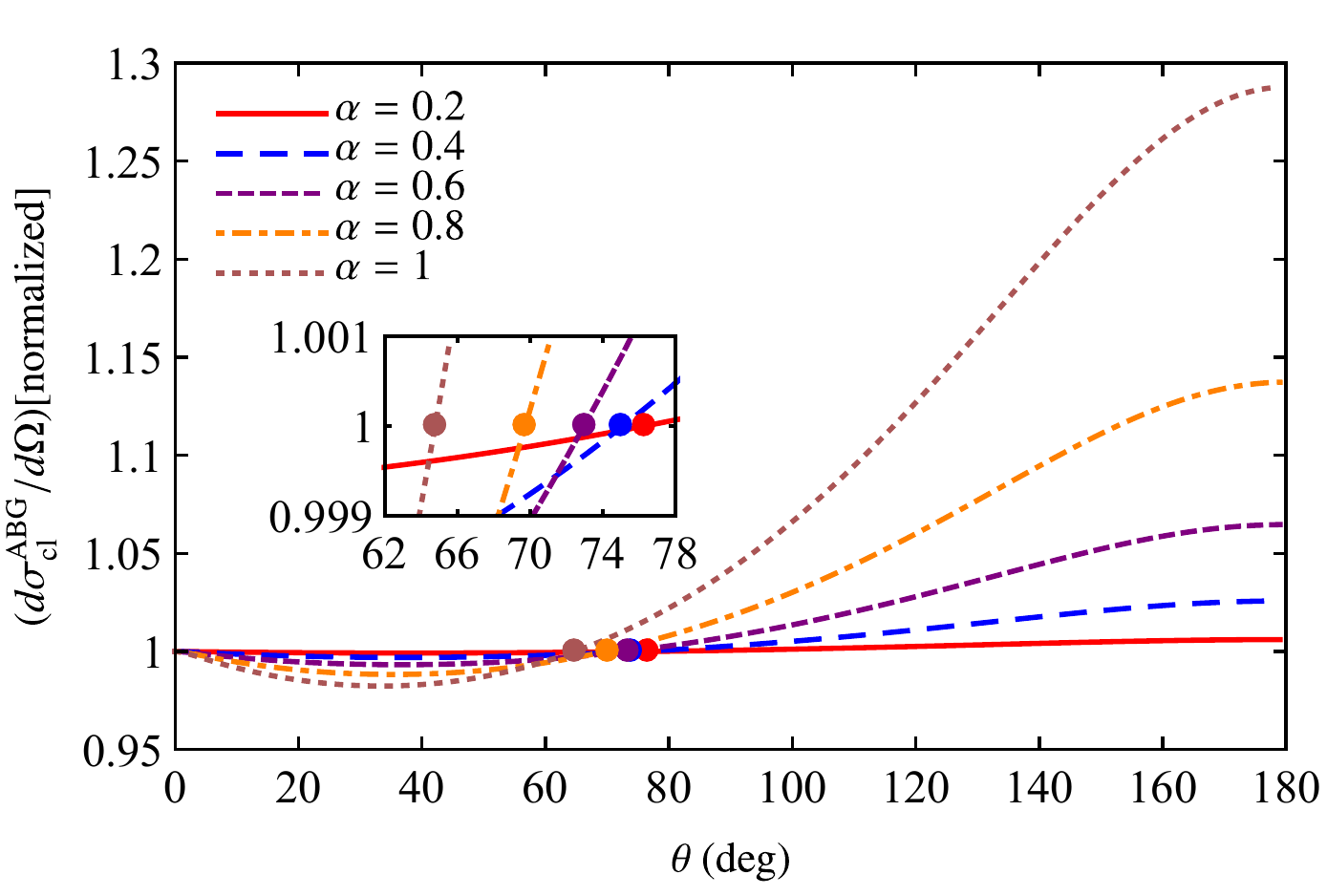}
    \caption{Classical differential SCS of the ABG RBH normalized by the Schwarzschild result, for different values of $\alpha$. The colored disks in the horizontal axis indicate the threshold values $\theta_{0}$, for which, when $\theta < \theta_{0}$, the corresponding differential SCS is smaller than the Schwarzschild one. For $\alpha = (0.2,0.4,0.6,0.8,1)$ we have $\theta_{0} = (76.25,75.028,72.969,69.733,64.774)$, respectively. The inset shows the plots near the values $\theta_{0}$}
    \label{CSCSABG}
\end{centering}
\end{figure}

\begin{figure}[!htbp]
\begin{centering}
    \includegraphics[width=\columnwidth]{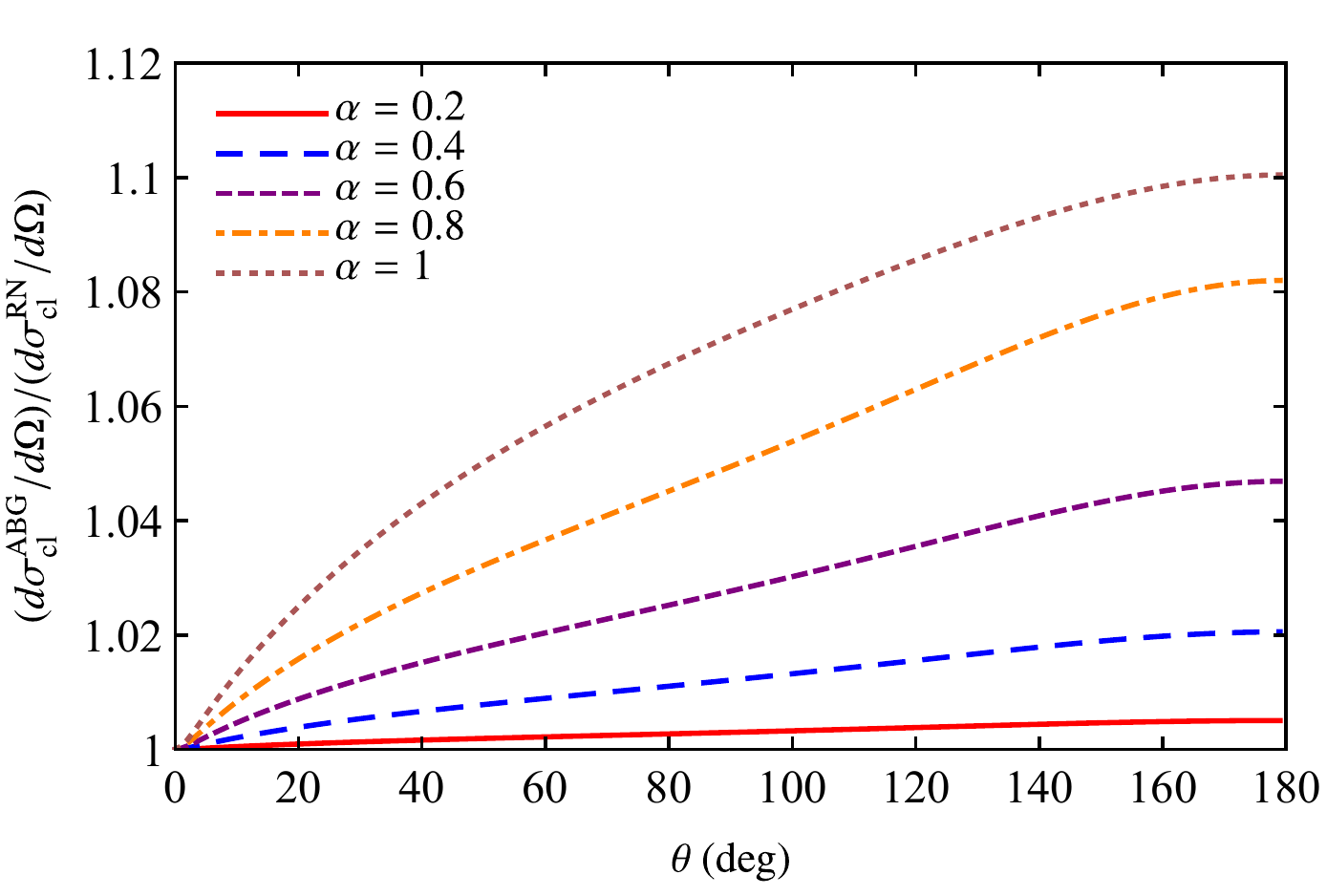}
    \caption{Ratio between the classical differential SCS of the ABG RBH and of the RN BH, for selected values of $\alpha$, as a function of the scattering angle $\theta$}
    \label{CSCSABGRN}
\end{centering}
\end{figure}

In Figs.~\ref{CSCSABG} and~\ref{CSCSABGRN} we show the classical differential SCS of the ABG RBH ($d\sigma^{\rm{ABG}}_{\rm{cl}} / d\Omega$), compared to the classical differential SCSs of Schwarzschild ($d\sigma^{\rm{S}}_{\rm{cl}} / d\Omega$) and RN ($d\sigma^{\rm{RN}}_{\rm{cl}} / d\Omega$) BHs, respectively, considering different choices of $\alpha$. In Fig.~\ref{CSCSABG} we see that for $\theta \lesssim 77^\circ$, the classical differential SCS of the ABG RBH can be smaller than the Schwarzschild case. This feature contrasts with the Bardeen case~\cite{MOC2015}, for which the classical differential SCS is bigger than the Schwarzschild one in the whole scattering angle range. We also note that the scattered flux near the backward direction ($\theta=\pi$) becomes more enhanced (with respect to the Schwarzschild case) for larger values of the normalized charge. 
Moreover, the classical differential SCS of the ABG RBH is typically bigger than that of the RN BH with the same value of $\alpha$, as shown in Fig.~\ref{CSCSABGRN}. In Sect.~\ref{sec:rst} we compare results obtained from Eq.~\eqref{CSCS} with those computed via the partial-waves method.

\subsection{Glory Approximation}\label{subsec:ga}
\begin{figure}[!htbp]
\begin{centering}
    \includegraphics[width=\columnwidth]{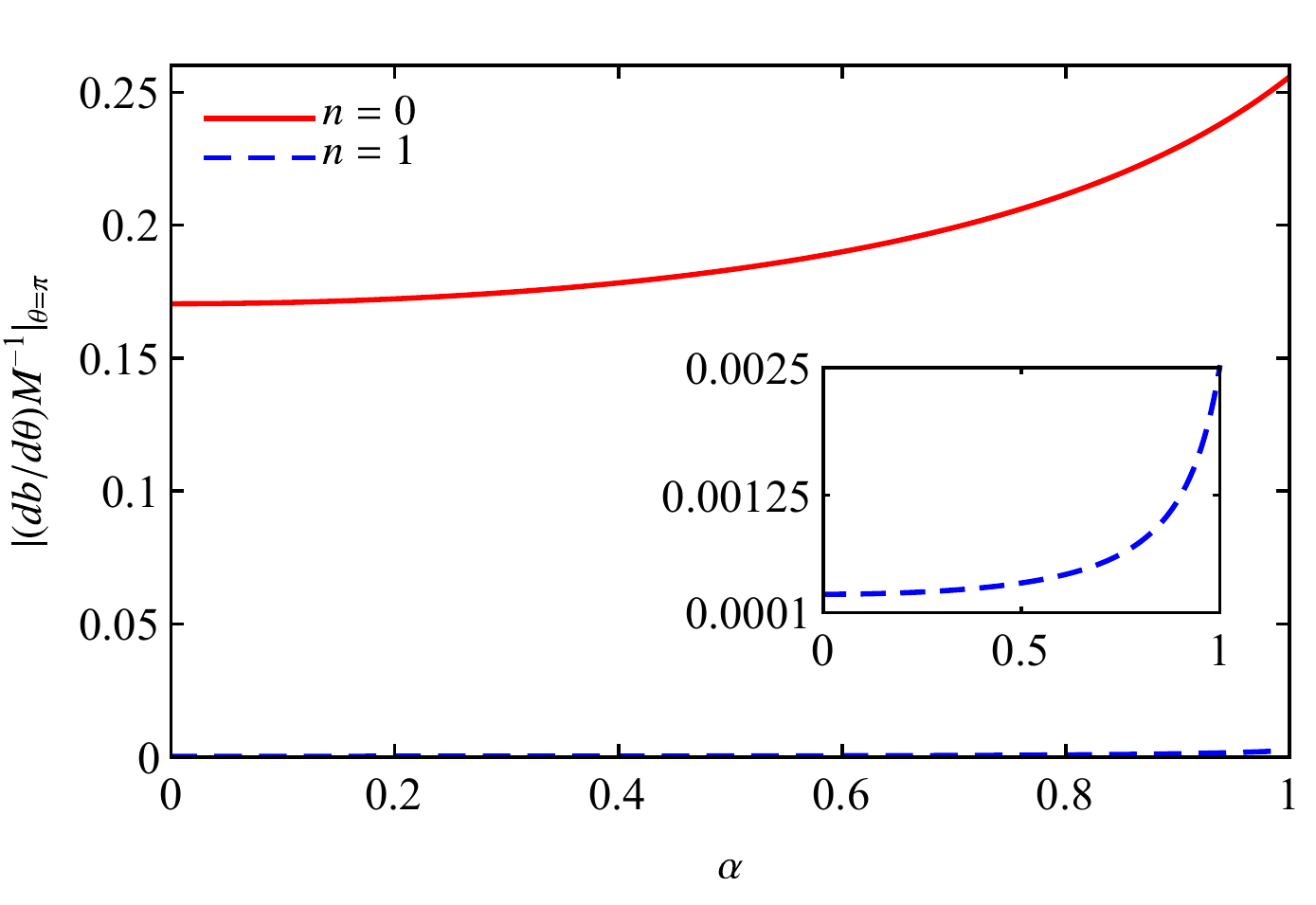}
    \caption{The quantity $\big|db/d\theta\big|_{\theta = \pi}$, as a function of $\alpha$, for $n = 0$ (solid red line) and $n = 1$ (dashed blue line) modes. In the inset, the result for $n = 1$ is emphasized}
    \label{GFD}
\end{centering}
\end{figure}

As a useful tool to anticipate wave interference effects of the scattering process near the backward direction, stands out the glory approximation. The differential SCS of a scalar wave in a spherically symmetric BH background can be written as~\cite{RAM1985}
\begin{equation}
\label{glory}\dfrac{d\sigma_{\rm{g}}}{d\Omega} = 2\pi \omega b_{g}^{2}\bigg|\dfrac{db}{d\theta}\bigg|_{\theta = \pi}J_{0}^{2}(\omega b_{g}\sin\theta),
\end{equation}
where $\omega$ is the frequency of the scalar wave, $b_{g}$ is the impact parameter of backscattered rays and $J_{0}$ is the Bessel function of the first kind. 
Notice that the glory scattering formula~\eqref{glory} is a semiclassical approximation, depending on both null geodesic and scalar wave parameters. This approximation is valid in the high-frequency limit, i.e., for $\omega M \gg 1$, although it also provides a good approximation for $\omega M \sim 1$. It is worth emphasizing that $b_{g}$ admits multiple values, which are related to the deflection angle, i.e., $\Theta = \pi + 2 n \pi$. We note that the $n = 0$ mode gives the largest contribution for the glory scattering. In Tab.~\ref{Tab.1} we show how much the glory intensity of the $n = 1$ mode corresponds to that of the $n = 0$ mode. Notice that the $n = 1$ mode has an intensity about less than $1\%$ of the $n = 0$ mode. As pointed in Ref.~\cite{MOC2015}, this is associated with the term $\big|db/d\theta\big|_{\theta = \pi}$ of the glory formula~\eqref{glory}, which is quickly suppressed as $n$ increases. To illustrated this, in Fig.~\ref{GFD} we present the behavior of $\big|db/d\theta\big|_{\theta = \pi}$ for $n = 0$ and $n = 1$ modes. Therefore, we will only consider the contributions of the $n = 0$ mode when using the semiclassical formula given by Eq.~\eqref{glory}.
\begin{table}[!htbp]
\centering
\begin{tabular}{||c|c|c|c|c|c|c||} 
 \hline
 $\alpha^{\rm{ABG}}$ & $0$ & $0.2$ & $0.4$ & $0.6$ & $0.8$ & $1$ \\ [0.5ex] 
 \hline
 $\%$ & $0.155$ & $0.16$ & $0.18$ & $0.226$ & $0.342$ & $0.848$  \\ \hline
\end{tabular}
\caption{{\footnotesize Percentual contribution to the glory intensity of the $n = 1$ mode, with respect to the $n = 0$ mode, considering different values of $\alpha$}}
\label{Tab.1}
\end{table} 

\begin{figure}[!htbp]
\begin{centering}
    \includegraphics[width=\columnwidth]{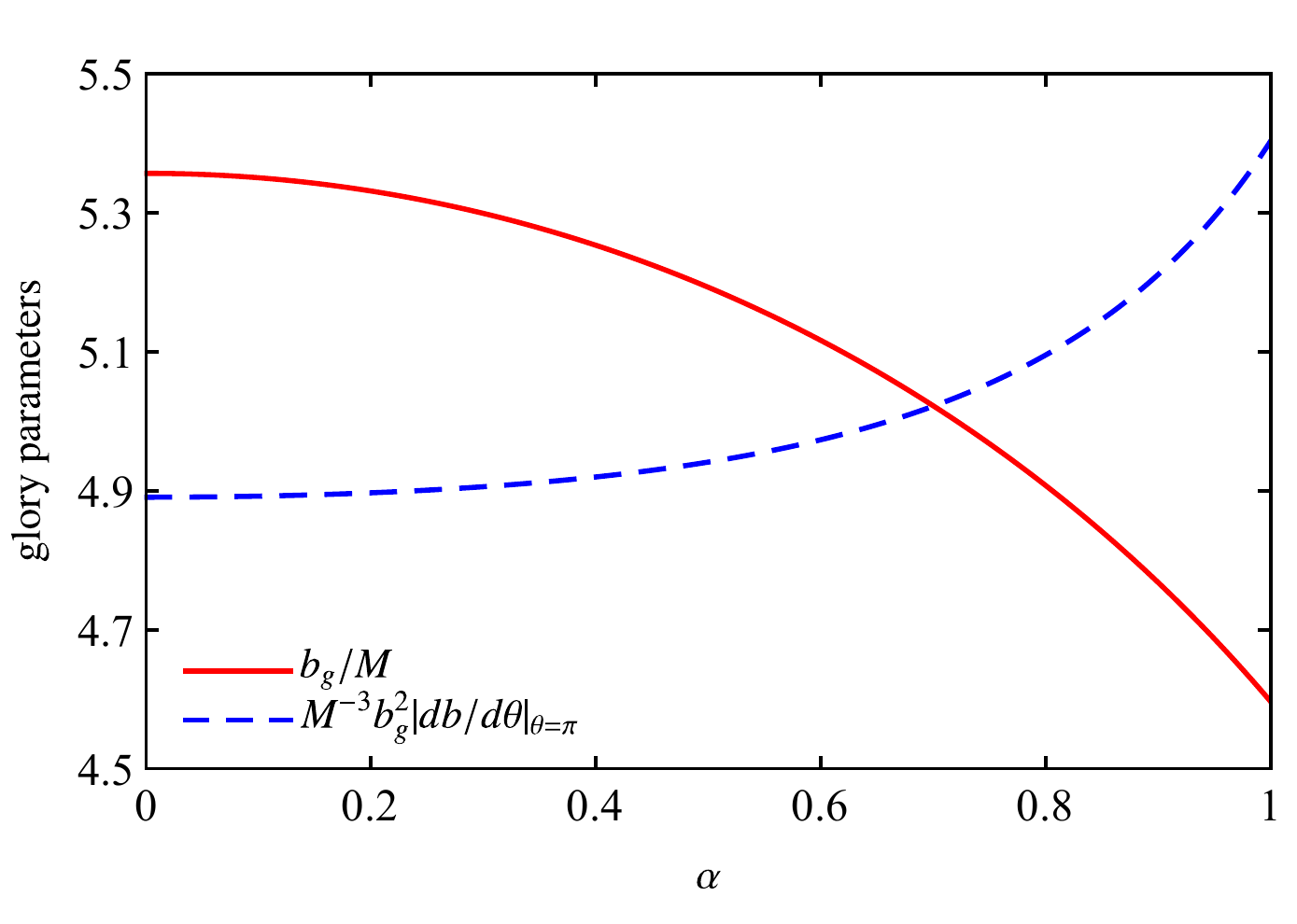}
    \caption{Glory parameters $b_{g}$ and $b_ {g}^{2}|db/d\theta|_{\theta = \pi}$, as functions of $\alpha$}
    \label{gp}
\end{centering}
\end{figure}

In Fig.~\ref{gp} we exhibit the glory parameters, namely $b_{g}$ and $b_ {g}^{2}|db/d\theta|_{\theta = \pi}$, as functions of the normalized electric charge, $\alpha$. We note that, as we increase the values of $\alpha$, $b_ {g}^{2}|db/d\theta|_{\theta = \pi}$ enhances, while $b_{g}$ diminishes. From their behavior we may anticipate two properties of the SCS, in the regime of validity of the glory approximation. The first one is the fact that the interference fringes get wider, as one increases the value of the BH charge, since the width is inversely proportional to $b_{g}$. The second one is that the backscattered flux intensity increases as we consider higher values of the BH charge, since it is proportional to $b_ {g}^{2}|db/d\theta|_{\theta = \pi}$. A similar behavior occurs for other BH spacetimes~\cite{CDE2009, MOC2015}. Note, however, that for the RN case $b_ {g}^{2}|db/d\theta|_{\theta = \pi}$ does not increase monotonically with the BH charge. Actually, it has a local minimum (cf. Fig. 10 of Ref.~\cite{CDE2009}). For its turn, when considering the Bardeen case, $b_ {g}^{2}|db/d\theta|_{\theta = \pi}$ increases monotonically with the BH charge (cf. Figs. 3 and 6 of Ref.~\cite{MOC2015}). Hence, we see that the glory scattering amplitude for some NED-based RBHs increases monotonically with the BH's charge, in contrast with the RN case.

\section{Partial-waves approach}\label{sec:ws}

\subsection{Scalar Field}\label{sec:sf}
\begin{figure}[!htbp]
\begin{centering}
    \includegraphics[width=\columnwidth]{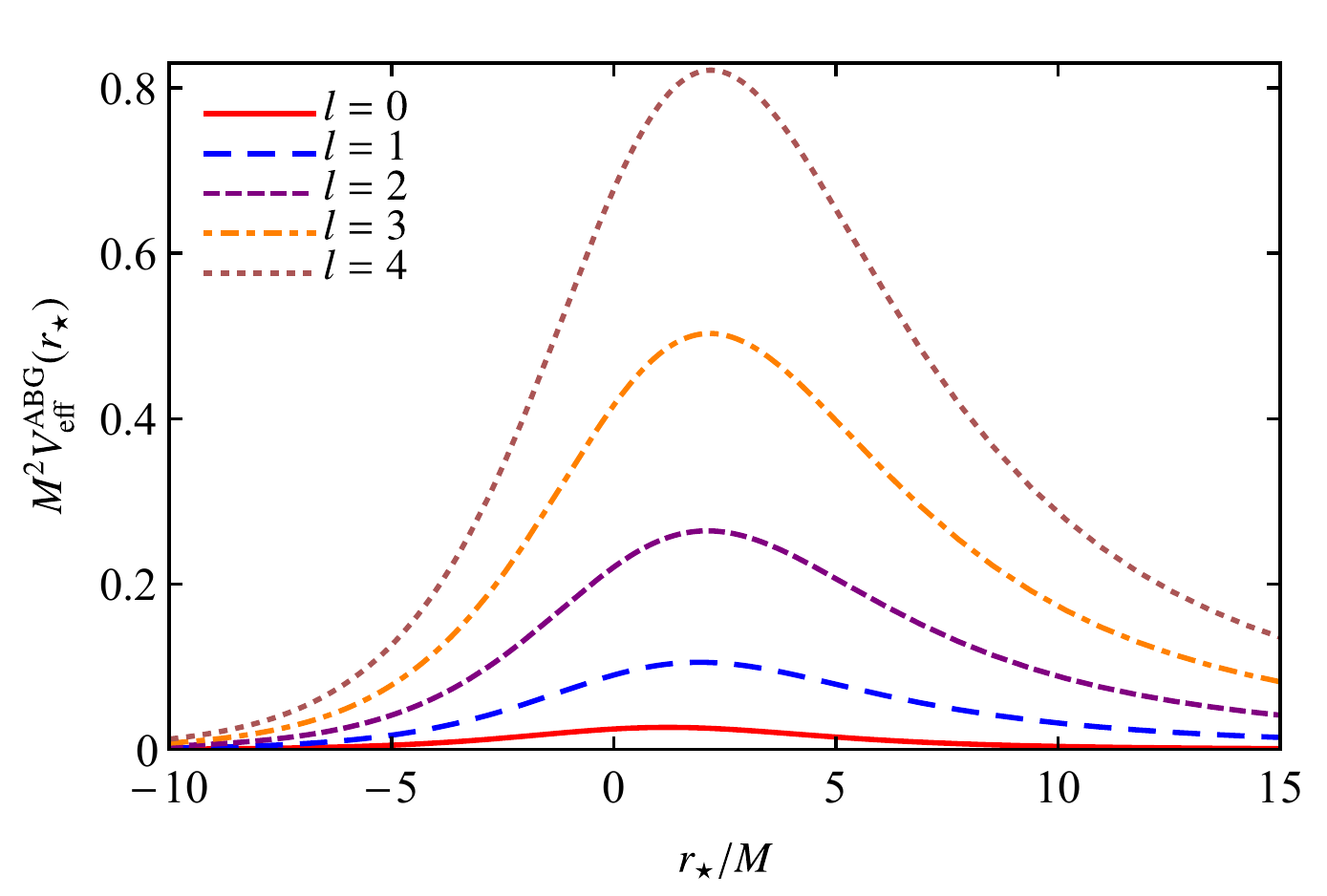}
    \caption{Effective potential of ABG RBHs, as a function of $r_{\star}$, for distinct choices of $l$, considering $\alpha = 0.5$}
    \label{ep1}
\end{centering}
\end{figure}

The Klein-Gordon (KG) equation governs the propagation of a massless and chargeless test scalar field $\Phi$. In the background of curved spacetimes, it can be written as
\begin{equation}
\label{KG}\nabla_{\mu}\nabla^{\mu}\Phi = \dfrac{1}{\sqrt{-g}}\partial_{\mu}\left(\sqrt{-g}g^{\mu \nu}\partial_{\nu}\Phi\right) = 0.
\end{equation}
Taking into account the spherical symmetry of the spacetime under consideration (cf. Eq.~\eqref{LE}), we can decompose the scalar field $\Phi$ as
\begin{equation}
\label{PHI}\Phi \equiv \sum_{l}^{\infty} C_{\omega l}\Phi_{\omega l} = \sum_{l}^{\infty} C_{\omega l}\frac{\Psi_{\omega l}(r)}{r}P_{l}(\cos\theta)e^{-i\omega t},
\end{equation}
where $\omega$ and $l$ are the frequency and the angular momentum of the scalar field mode, respectively, and the constant coefficients $C_{\omega l}$ are determined by the boundary conditions. The functions $P_{l}(\cos\theta)$ are the Legendre polynomials and $\Psi_{\omega l}(r)$ are radial functions. Using the tortoise coordinate $r_{\star}$, which can be obtained from the relation $dr=f(r)dr_{\star}$, we can show that $\Psi_{\omega l}(r)$ satisfies 
\begin{equation}
\label{RE_KG}\frac{d^{2}}{dr_{\star}^{2}}\Psi_{\omega l}+\left(\omega^{2}-V_{\rm{eff}}(r)\right)\Psi_{\omega l}=0,
\end{equation}
where the effective potential $V_{\rm{eff}}(r)$ is
\begin{equation}
\label{EP}V_{\rm{eff}}(r) \equiv f(r)\left(\dfrac{1}{r}\dfrac{df(r)}{dr}+\dfrac{l(l+1)}{r^{2}}\right).
\end{equation}
Notice that the domain of the tortoise coordinate is $(-\infty,\infty)$, whereas the domain of the coordinate $r$ is $[r_{+},\infty)$. In Fig.~\ref{ep1} we show the effective potential of the ABG RBH, $V^{\rm{ABG}}_{\rm{eff}}$, for different values of $l$. We see that the potential barrier grows as we increase $l$ and also presents the following asymptotic limits:
\begin{equation}
\label{Asym_EP}\lim_{r_{\star} \rightarrow \pm \infty} V_{\rm{eff}}(r_{\star}) = 0.
\end{equation}

\begin{figure}[!htbp]
\begin{centering}
    \includegraphics[width=\columnwidth]{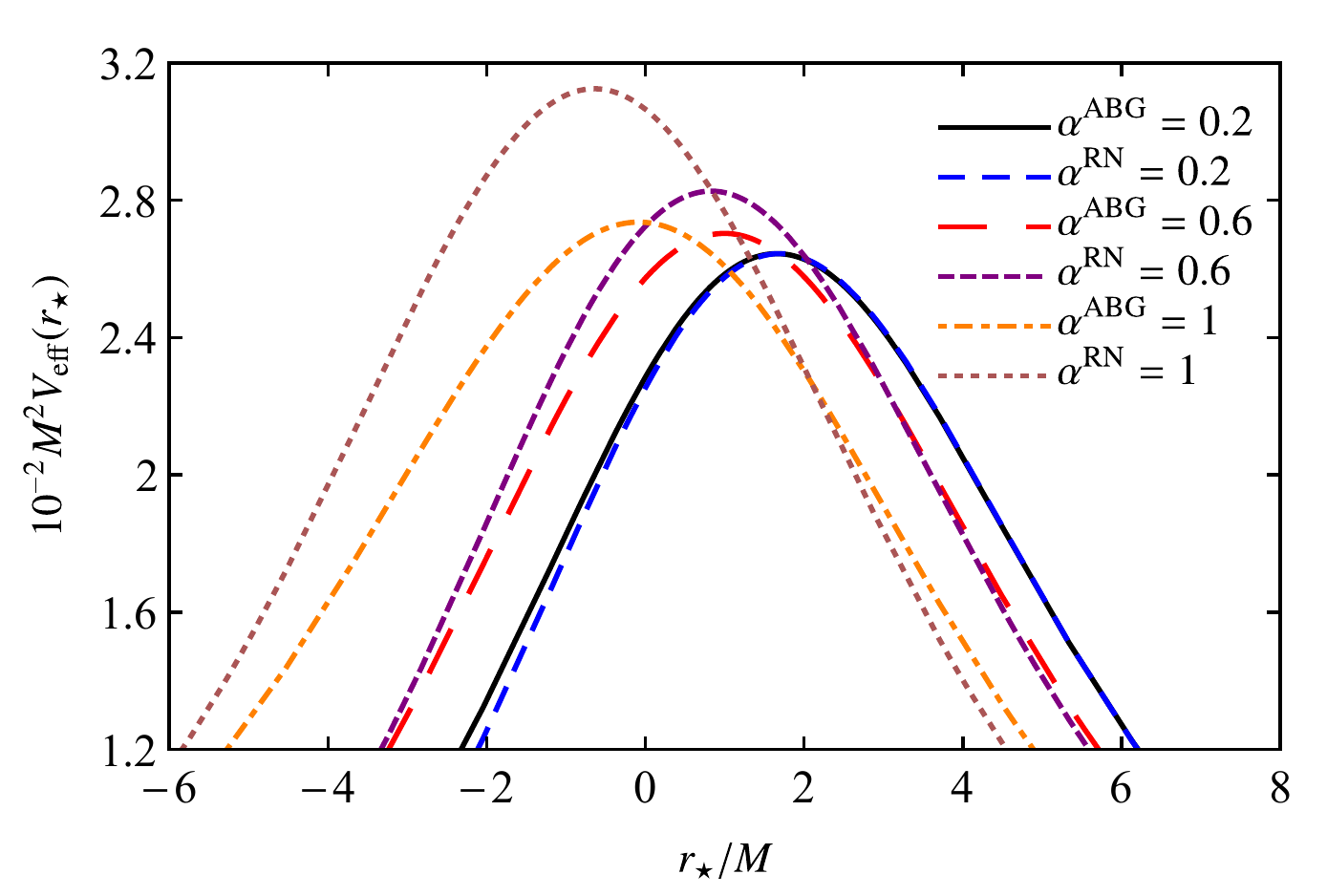}
    \caption{The behavior of the effective potentials of ABG and RN BHs, as functions of $r_{\star}$, near their local maxima, for different values of $\alpha$ and with $l = 0$}
    \label{ep2}
\end{centering}
\end{figure}

In Fig.~\ref{ep2} we compare the local maximum of the effective potentials of ABG and RN BHs, for distinct values of $\alpha$. In general, the height of the peaks increases as we consider higher values of $\alpha$, with those of the ABG RBH being typically smaller than the corresponding RN case (for the same value of $\alpha$). However, for small values of $\alpha$, the magnitude of the peaks of ABG and RN effective potentials are very similar. 

\subsection{Scattering Cross Section}\label{subsec:scs}
For the BH scattering problem, we assume the following boundary conditions
\begin{equation}
\label{BC}\Psi_{\omega l}\sim\begin{cases}
T_{\omega l}e^{-i\omega r_{\star}}, & r_{\star}\rightarrow -\infty,\\
e^{-i\omega r_{\star}}+R_{\omega l}e^{i\omega r_{\star}}, & r_{\star}\rightarrow \infty,
\end{cases}
\end{equation}
where $T_{\omega l}$ and $R_{\omega l}$ are complex coefficients. The solutions~\eqref{BC} are the so-called {\it{in modes}}, which correspond to plane waves purely incoming from the past null infinity. From the conservation of the flux, one can show that
\begin{equation}
\label{CF}|T_{\omega l}|^2 + |R_{\omega l}|^2 = 1,
\end{equation}
in which $|T_{\omega l}|^2$ and $|R_{\omega l}|^2$ are the transmission and reflection coefficients, respectively. 

The differential SCS for static and spherically spacetimes is given by~\cite{FHM1988}
\begin{equation}
\label{SCS}\dfrac{d\sigma}{d\Omega} = |h(\theta)|^{2},
\end{equation}
where the scattering amplitude $h(\theta)$ reads
\begin{equation}
\label{scatta}h(\theta) = \dfrac{1}{2i\omega}\sum_{l = 0}^{\infty}(2l+1)[e^{2i\delta_{l}(\omega)}-1]P_{l}(\cos\theta),
\end{equation}
with the phase shifts $e^{2i\delta_{l}(\omega)}$ defined as
\begin{equation}
\label{ps}e^{2i\delta_{l}(\omega)} \equiv (-1)^{l+1}R_{\omega l}.
\end{equation}

In order to compute Eq.~\eqref{SCS} via the partial-waves method, we need to obtain the coefficients $R_{\omega l}$. We have applied the stiffness switching integration method~\cite{WP2007} to solve Eq.~\eqref{RE_KG} numerically from near the event horizon, $r_{h}$, up to some radial position far from the BH. Then we match the numerical solutions of Eq.~\eqref{RE_KG} with the appropriated boundary conditions given by Eq.~\eqref{BC}. We also adopt the convergence method developed in Refs.~\cite{YRW1954,DDL2006} to improve the results obtained from Eq.~\eqref{scatta}. 

\section{Results}\label{sec:rst}

\subsection{Scalar Scattering by ABH RBH}
\begin{figure}[!htbp]
\begin{centering}
    \includegraphics[width=\columnwidth]{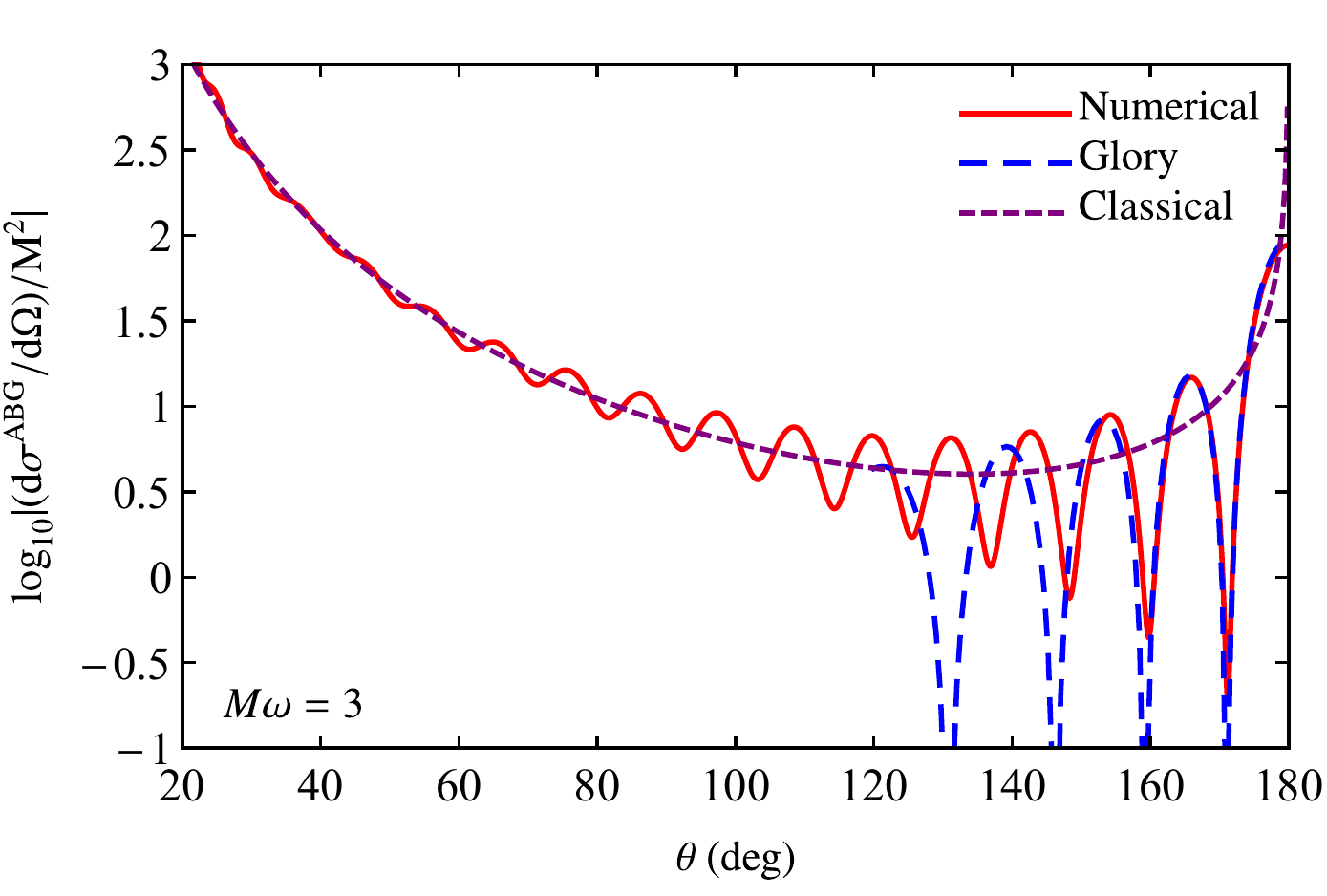}
    \includegraphics[width=\columnwidth]{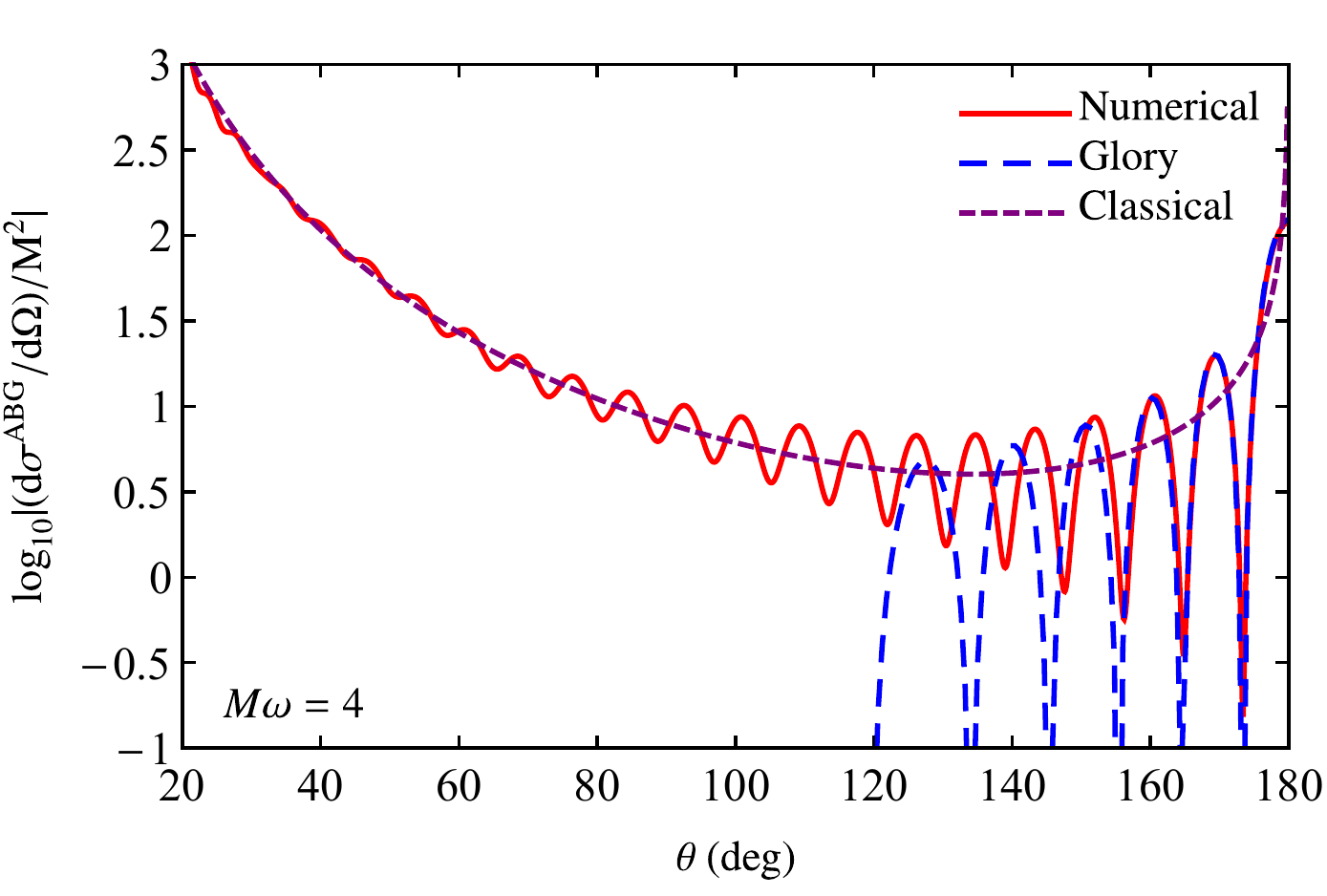}
    \caption{Differential SCS of the ABG RBH, with $\alpha = 0.5$, for: (i) $M\omega = 3$ (top panel) and (ii) $M\omega = 4$ (bottom panel). We also plot the corresponding classical and glory approximations in each case}
    \label{CSSCS}
\end{centering}
\end{figure}

In Fig.~\ref{CSSCS} we compare the differential SCS of massless scalar waves by ABG RBH, $d\sigma^{\rm{ABG}} / d\Omega$, computed numerically with the classical and semiclassical results. We note that, as a consequence of the weak-field deflection (see Eq.~\eqref{DA_WFL}), the differential SCS diverges as we approach the forward direction ($\theta\rightarrow 0^{\circ}$). As the backward direction ($\theta \rightarrow 180^{\circ}$) is approached, the numerical results oscillate around the classical ones, and they are well described in this regime by the glory approximation. In the semiclassical picture, the oscillations exhibited by the SCS can be interpreted as a consequence of the interference between waves orbiting the BH in opposite senses.

\begin{figure*}
\begin{centering}
    \includegraphics[width=\columnwidth]{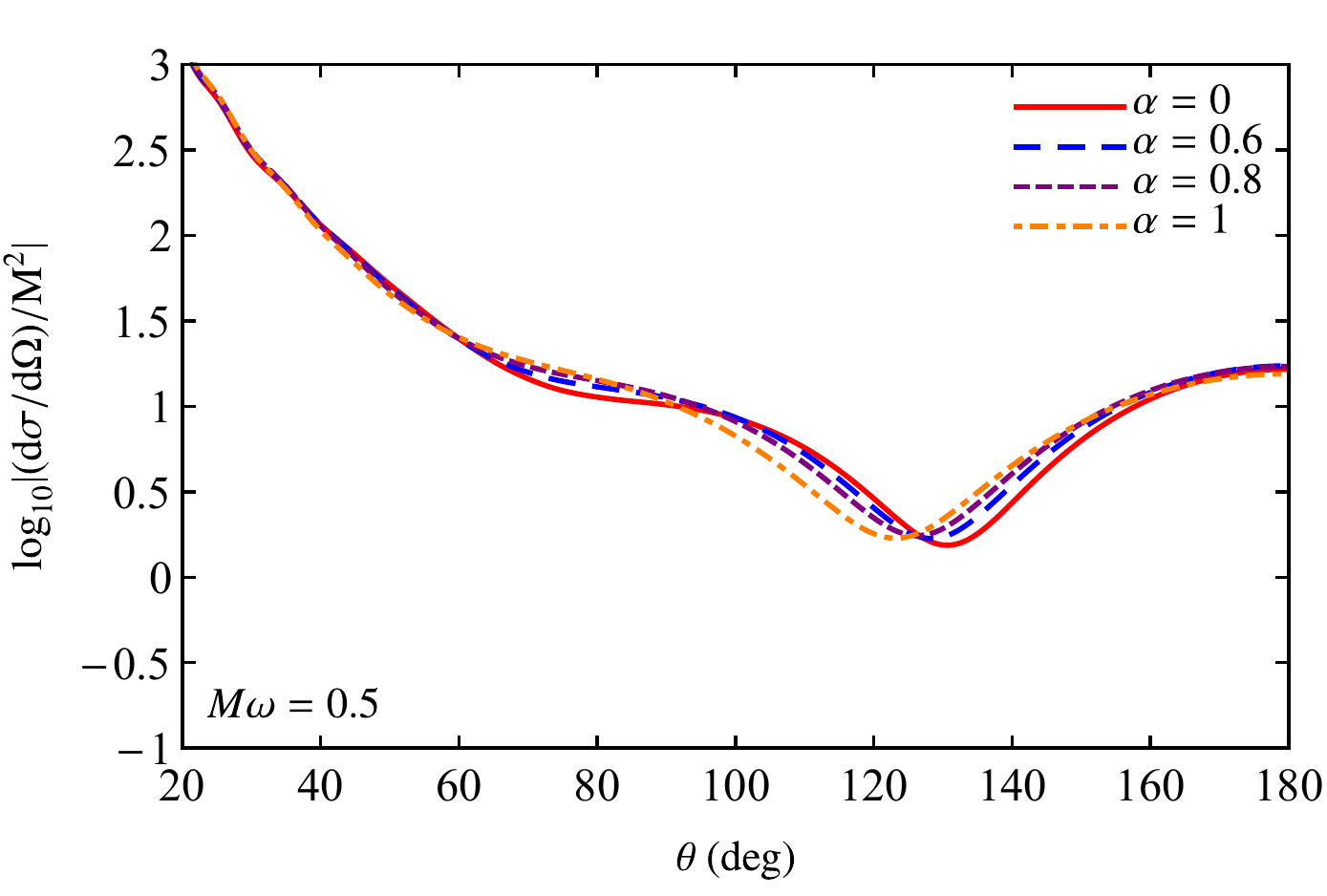}
    \includegraphics[width=\columnwidth]{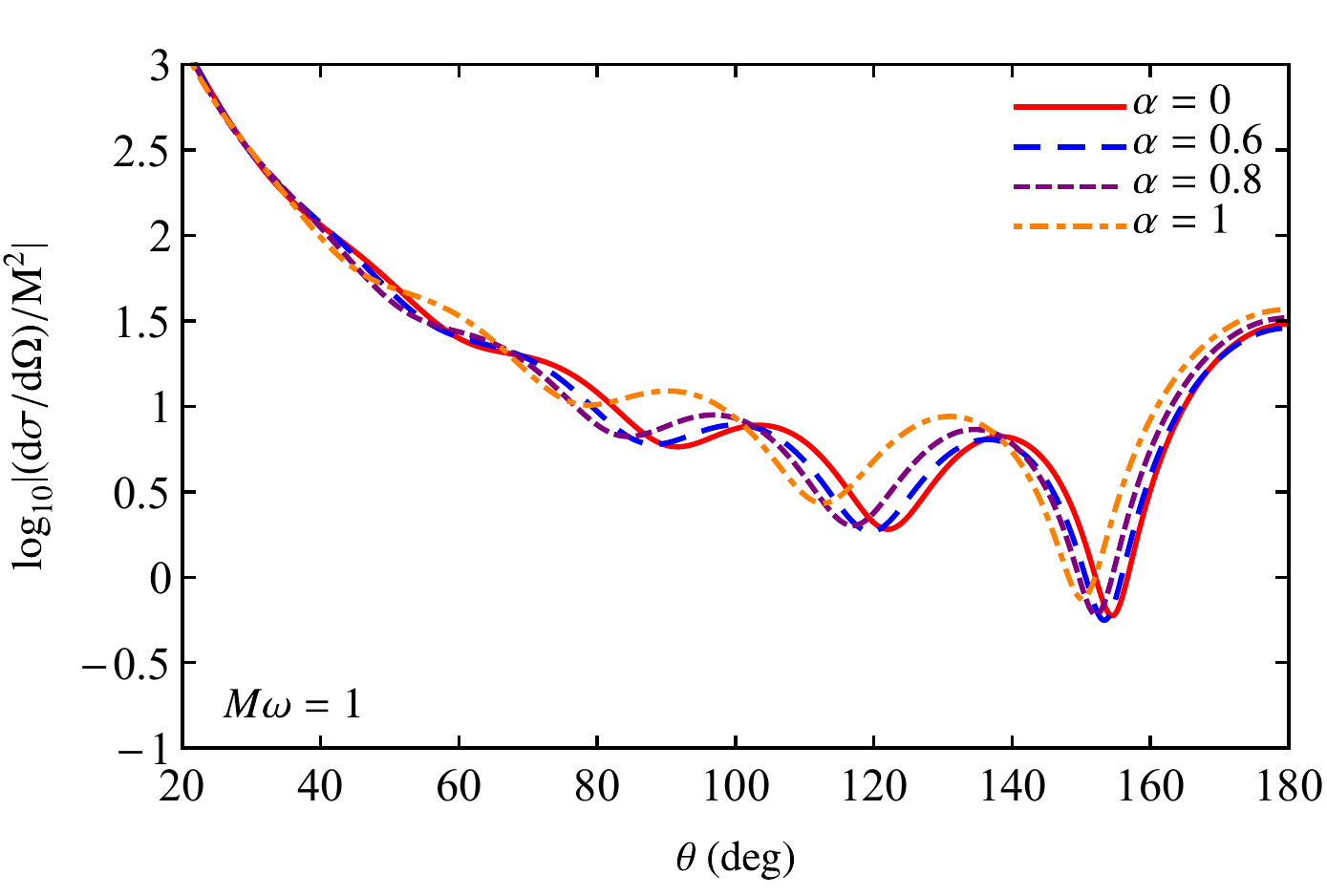}
    \includegraphics[width=\columnwidth]{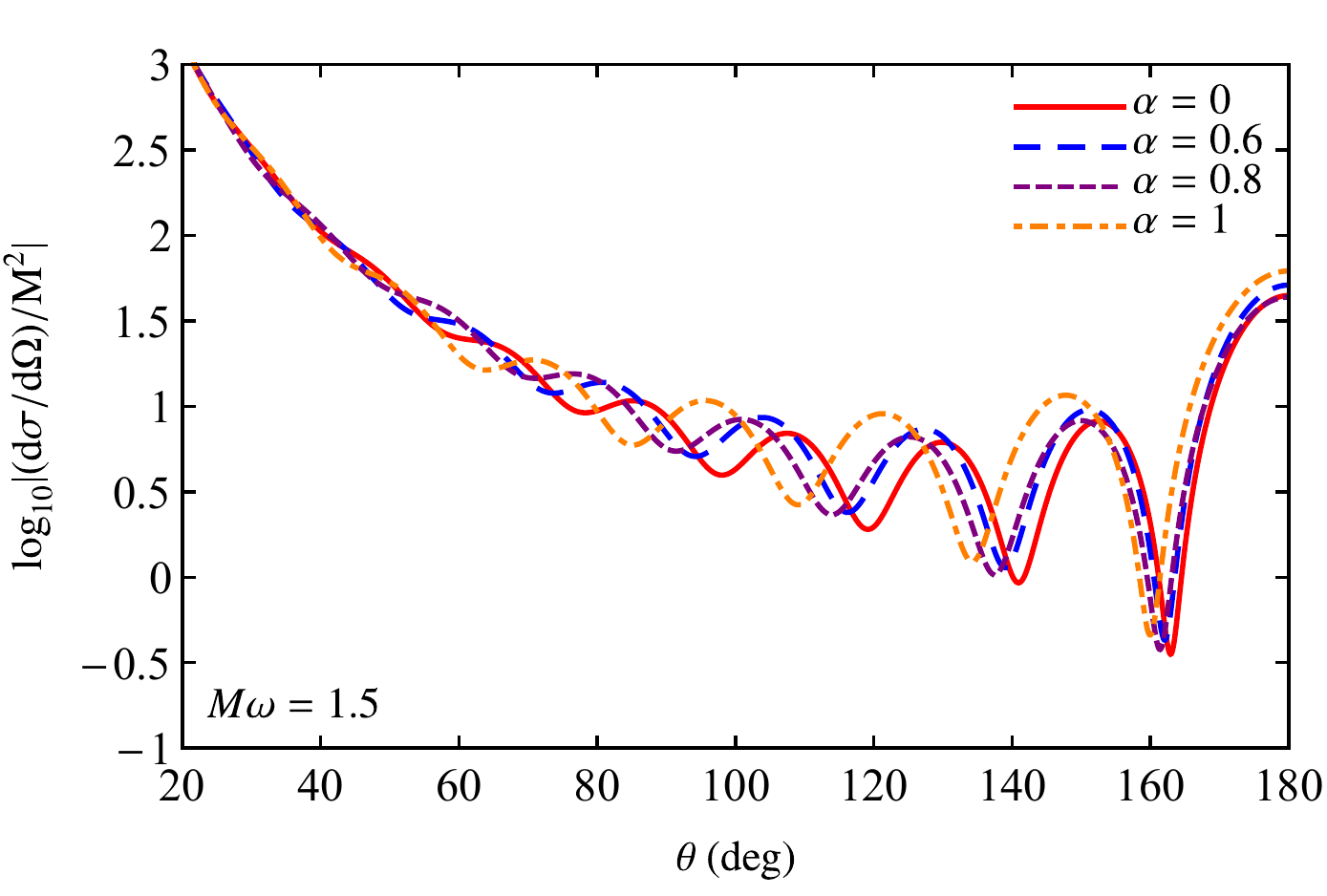}
    \includegraphics[width=\columnwidth]{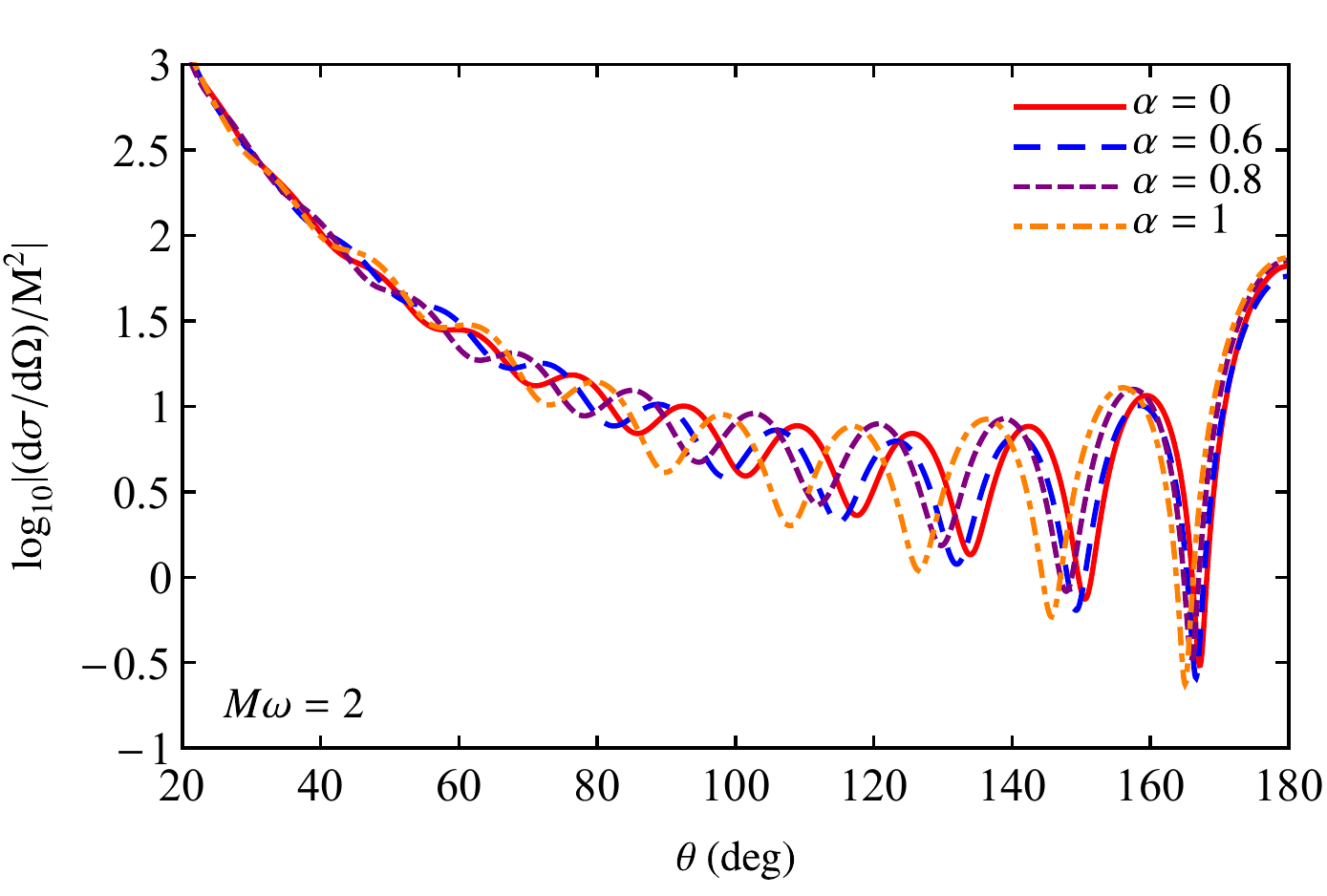}
    \caption{Differential SCS of ABG RBHs for distinct values of $\alpha$ and $\omega$. For comparison, we also show the Schwarzschild result ($\alpha = 0$)}
    \label{SCSs}
\end{centering}
\end{figure*}

In Fig.~\ref{SCSs} we plot the differential SCS of ABG RBH for different values of $\alpha$ and $\omega$, as well as for the Schwarzschild case ($\alpha = 0$), for comparison. We see that the interference fringe widths increase as we consider higher values of the normalized charge $\alpha$, for a fixed value of $ \omega$, or as we decrease the values of $\omega$, for a fixed value of $\alpha$. These features are in accordance with the semiclassical analysis (cf. Sect.~\ref{subsec:ga}). As anticipated from Eq.~\eqref{glory}, the interference fringes width are proportional to $1/(b_{g}\omega)$ and $b_{g}$ decreases monotonically with the increase of $\alpha$ (cf. Fig.~\ref{gp}). Moreover, we observe that, for small scattering angles, the behavior of the SCS is not substantially affected by the BH charge, as expected (cf. Sect.~\ref{subsec:gs}, and, in particular, Eqs.~\eqref{DA_WFL} and~\eqref{CSCS_SA}).

\begin{figure}[!htbp]
\begin{centering}
    \includegraphics[width=\columnwidth]{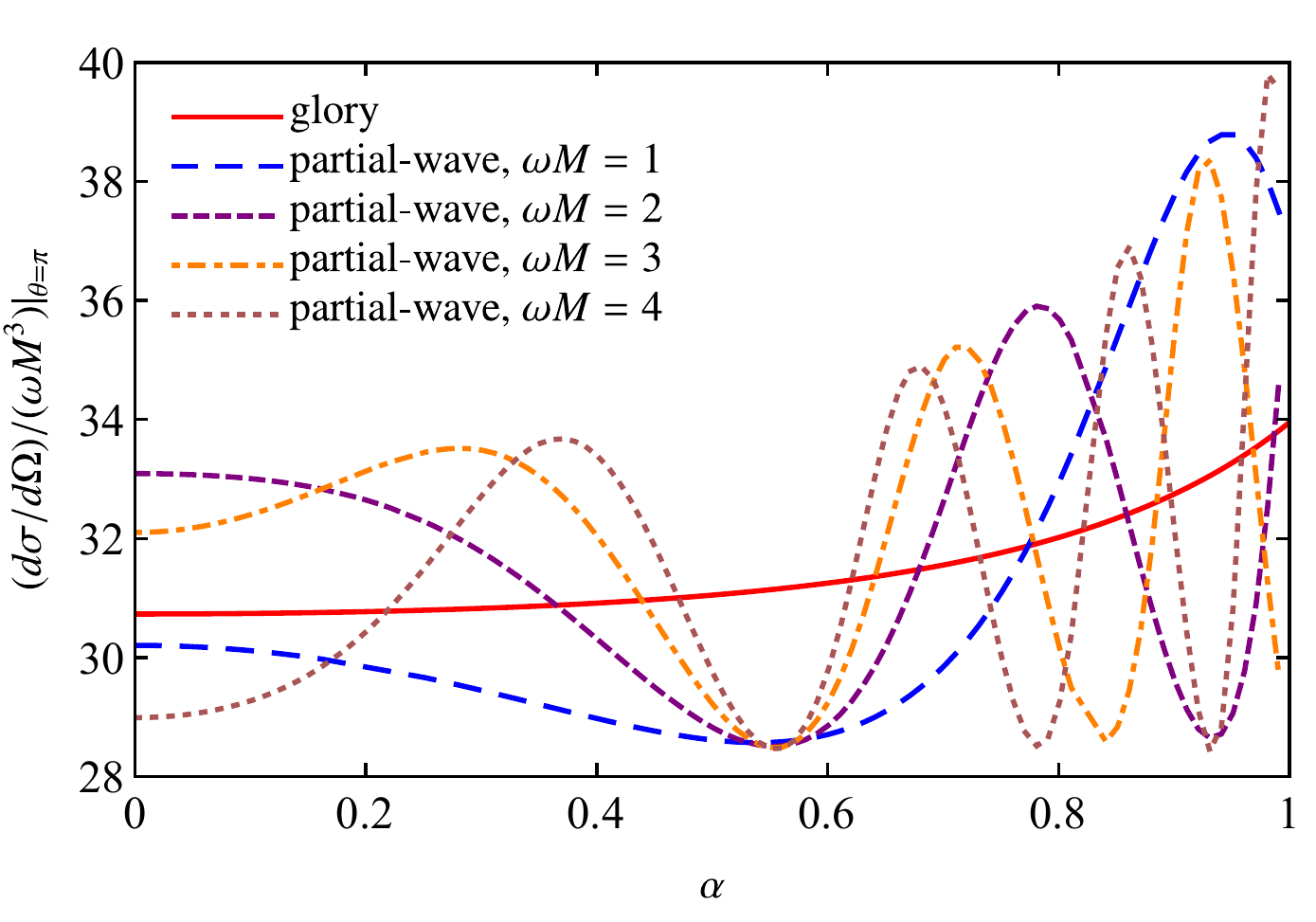}
    \caption{Intensity of the glory peak, as a function of $\alpha$, considering the glory approximation and the partial-waves method, for distinct choices of $M\omega$}
    \label{NGP}
\end{centering}
\end{figure}

In Fig.~\ref{NGP} we compare the amplitude of the backscattered wave computed via the partial-waves method (for different values of $\omega$) with the glory approximation. As we can see, the results obtained through the partial-waves method oscillate around the ones obtained using the semiclassical glory approximation. This behavior is also observed for RN~\cite{CDE2009}, Bardeen~\cite{MOC2015} and charged dilatonic~\cite{HZ2020} BHs.

\begin{figure}[!htbp]
\begin{centering}
    \includegraphics[width=\columnwidth]{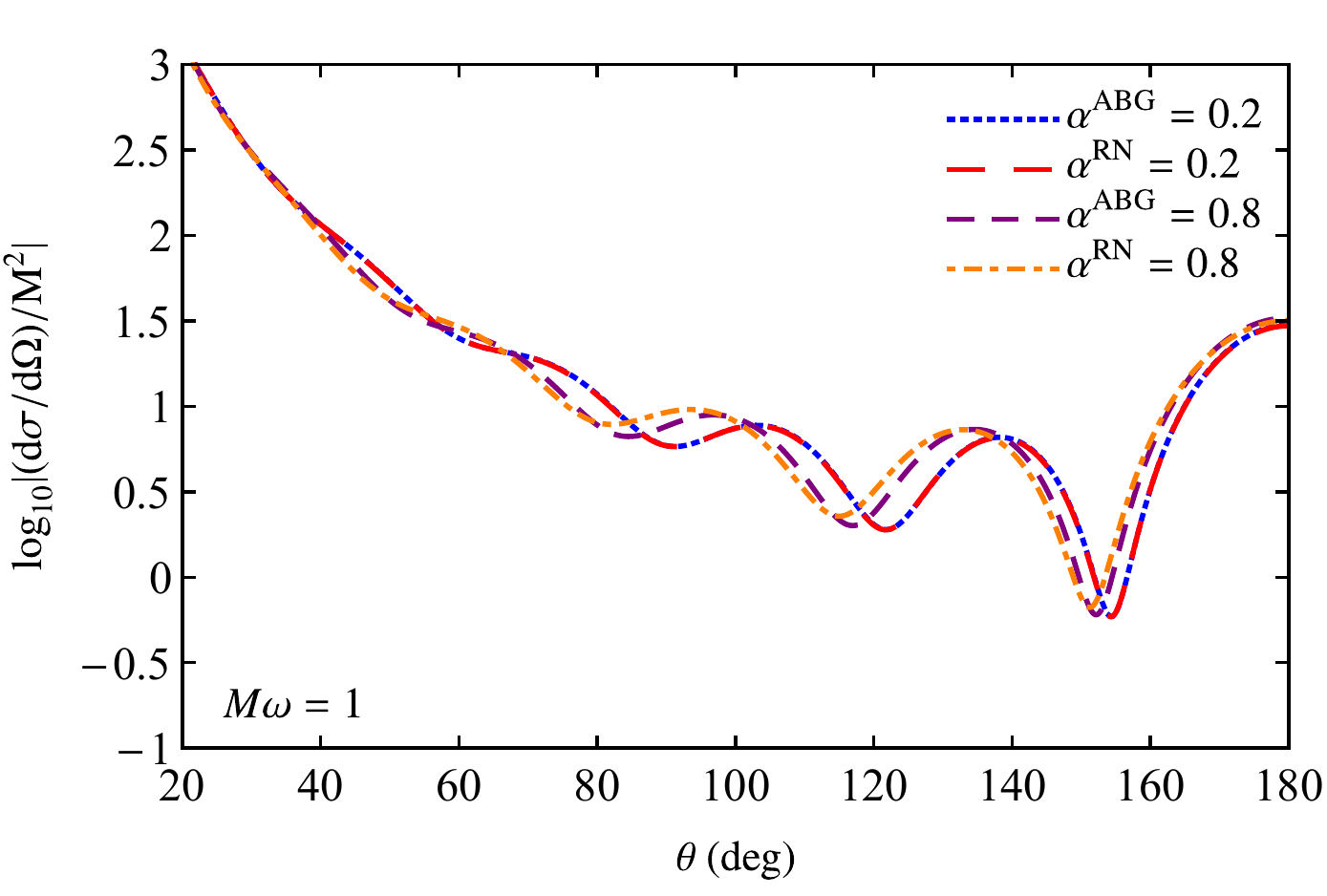}
    \caption{Differential SCS of ABG and RN BHs for two distinct values of $\alpha$, considering $M \omega = 1$ in both cases}
    \label{SCSABGRN}
\end{centering}
\end{figure}

A comparison between the SCSs of ABG and RN BHs for two fixed values of $\alpha$ is exhibit in Fig~\ref{SCSABGRN}. We note that, for $\alpha=0.2$, the ABG and RN BHs present a very similar scattering pattern for any value of the scattering angle, showing that, regarding the scattering of massless scalar waves, the two BH solutions can be indistinguishable. In the following Subsection, we analyze this similarity in more detail.

\subsection{Similar Scattering properties of ABG and RN BHs}
\begin{figure}[!htbp]
\begin{centering}
    \includegraphics[width=\columnwidth]{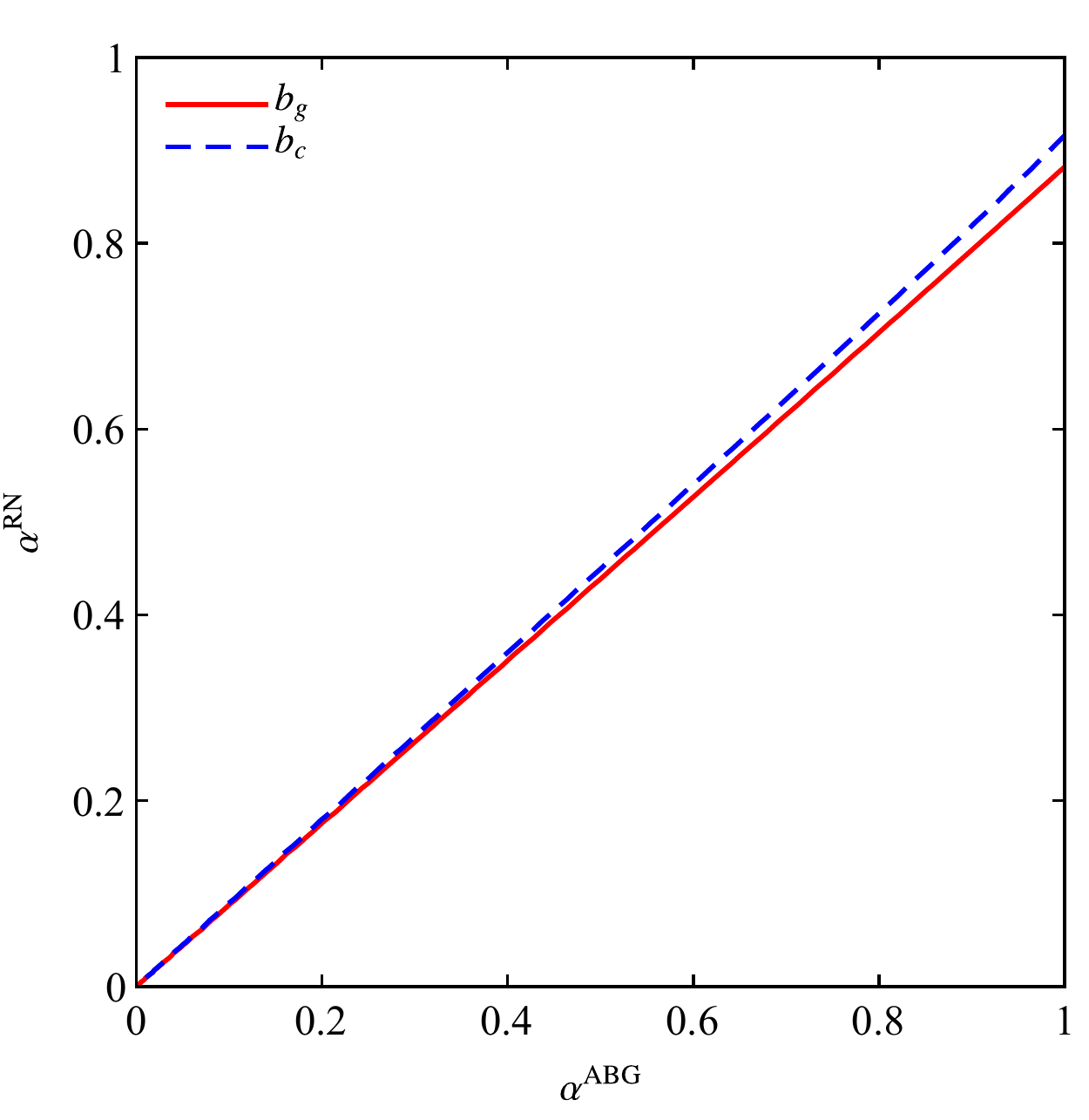}
    \caption{Values of the pair $(\alpha^{\rm{ABG}},\alpha^{\rm{RN}})$ for which their critical impact parameter~($b_c$, dashed blue line) and impact parameter of backscattered rays~($b_g$, solid red line) coincide}
    \label{ipbr}
\end{centering}
\end{figure}

In Ref.~\cite{PLC2020}, it was shown that ABG and RN BHs can have similar absorption properties, as long as we consider small-to-moderate values of the BH charge. To find BH configurations whose absorption cross sections were similar, we searched for values of the pair ($\alpha^{\rm{ABG}},\alpha^{\rm{RN}}$) that lead to the same geometric cross section, namely $\sigma_{\rm{geo}} = \pi b_{c}^{2}$~\cite{W1984}. Here, we follow a similar strategy, seeking for values of ($\alpha^{\rm{ABG}},\alpha^{\rm{RN}}$) which lead to the same value for the glory impact parameter, $b_{g}$. In Fig.~\ref{ipbr} we show the values of the normalized charge for which $b_{g}$ and $b_{c}$ assume the same value for ABG and RN BHs. We observe that $b_{c}$ and $b_{g}$ present a similar behavior, and we point out that the values of $b_{g}$ are found up to $(\alpha^{\rm{ABG}},\alpha^{\rm{RN}}) = (1,0.8825)$.

\begin{figure*}
\begin{centering}
    \includegraphics[width=\columnwidth]{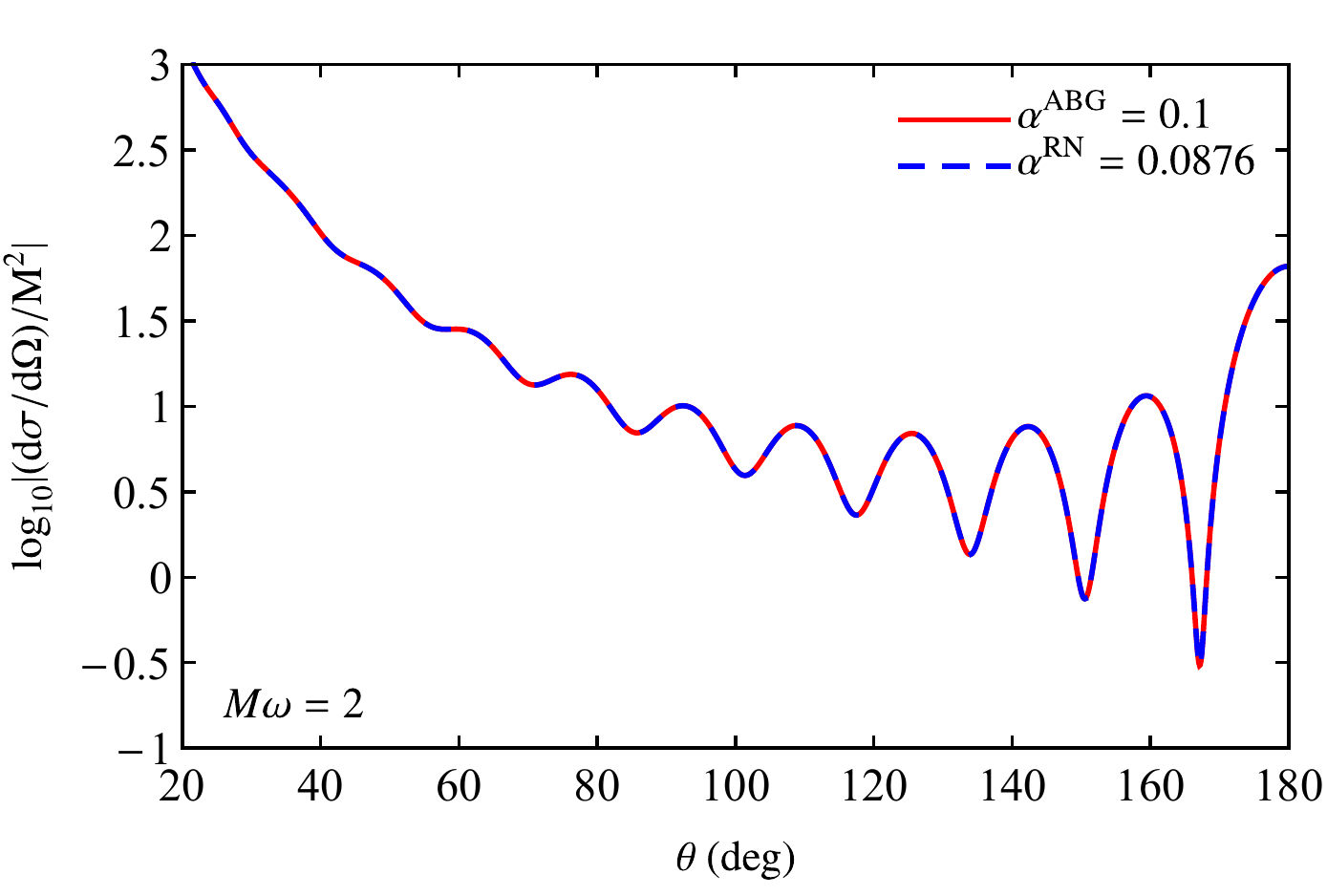}
    \includegraphics[width=\columnwidth]{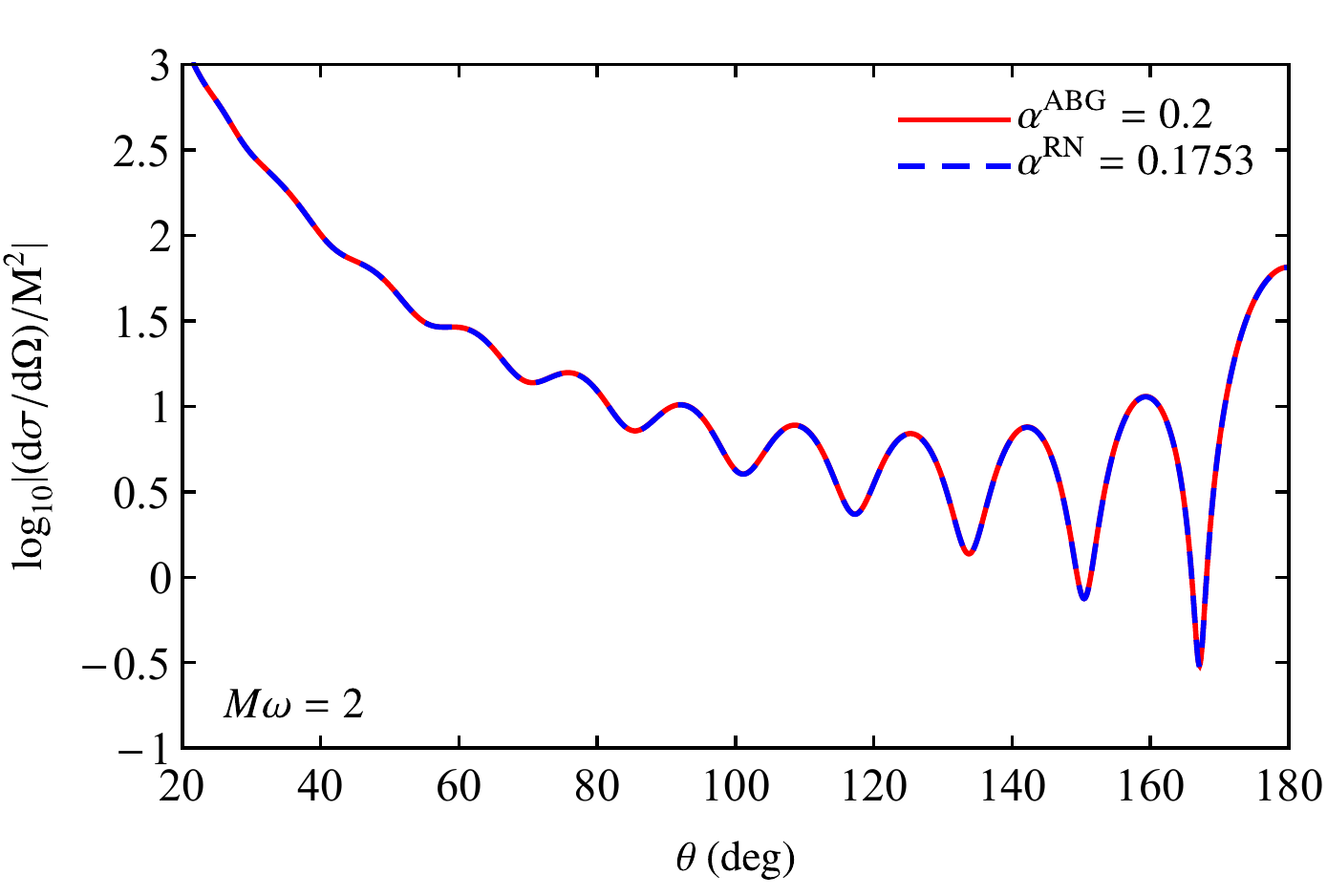}
    \includegraphics[width=\columnwidth]{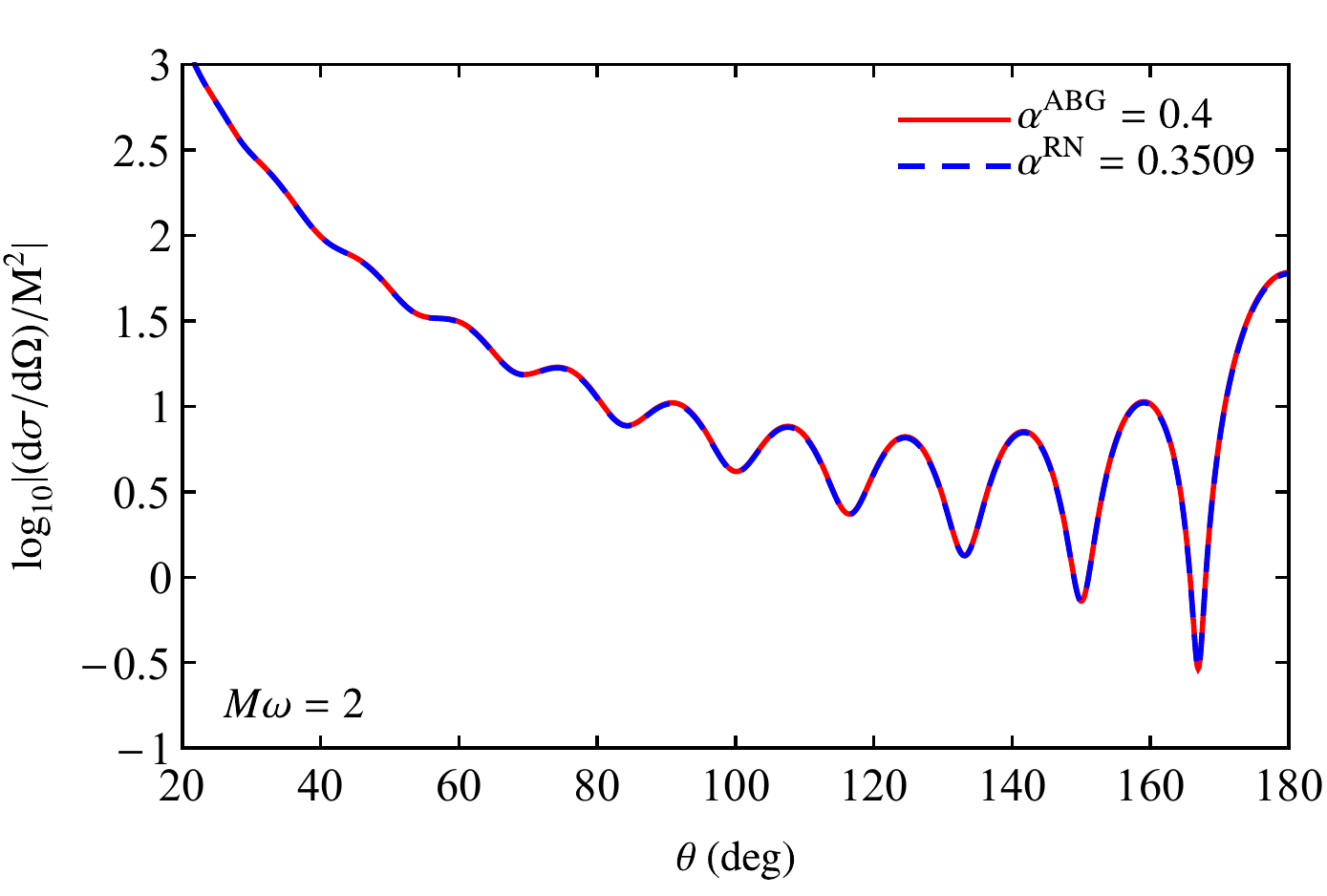}
    \includegraphics[width=\columnwidth]{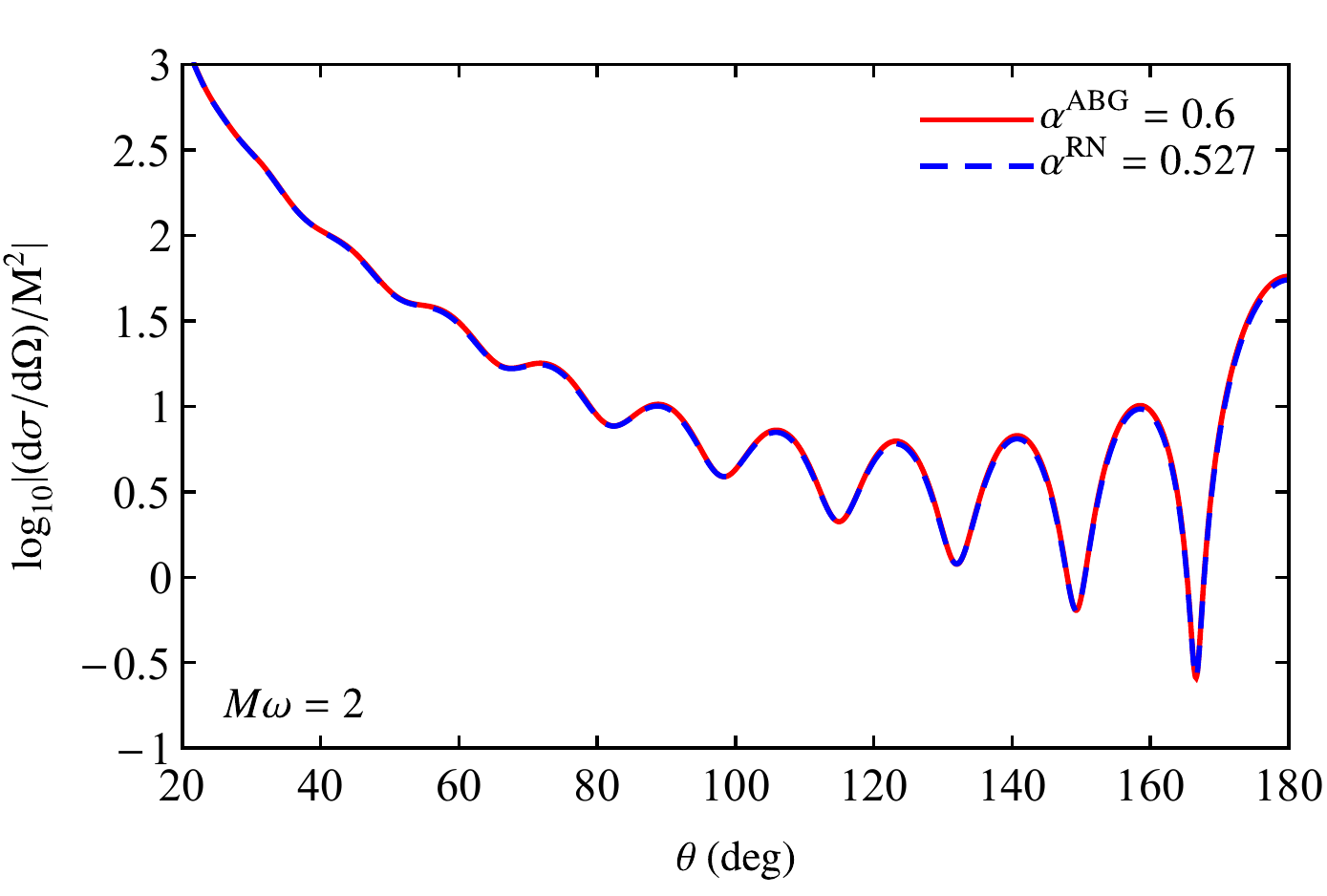}
	\includegraphics[width=\columnwidth]{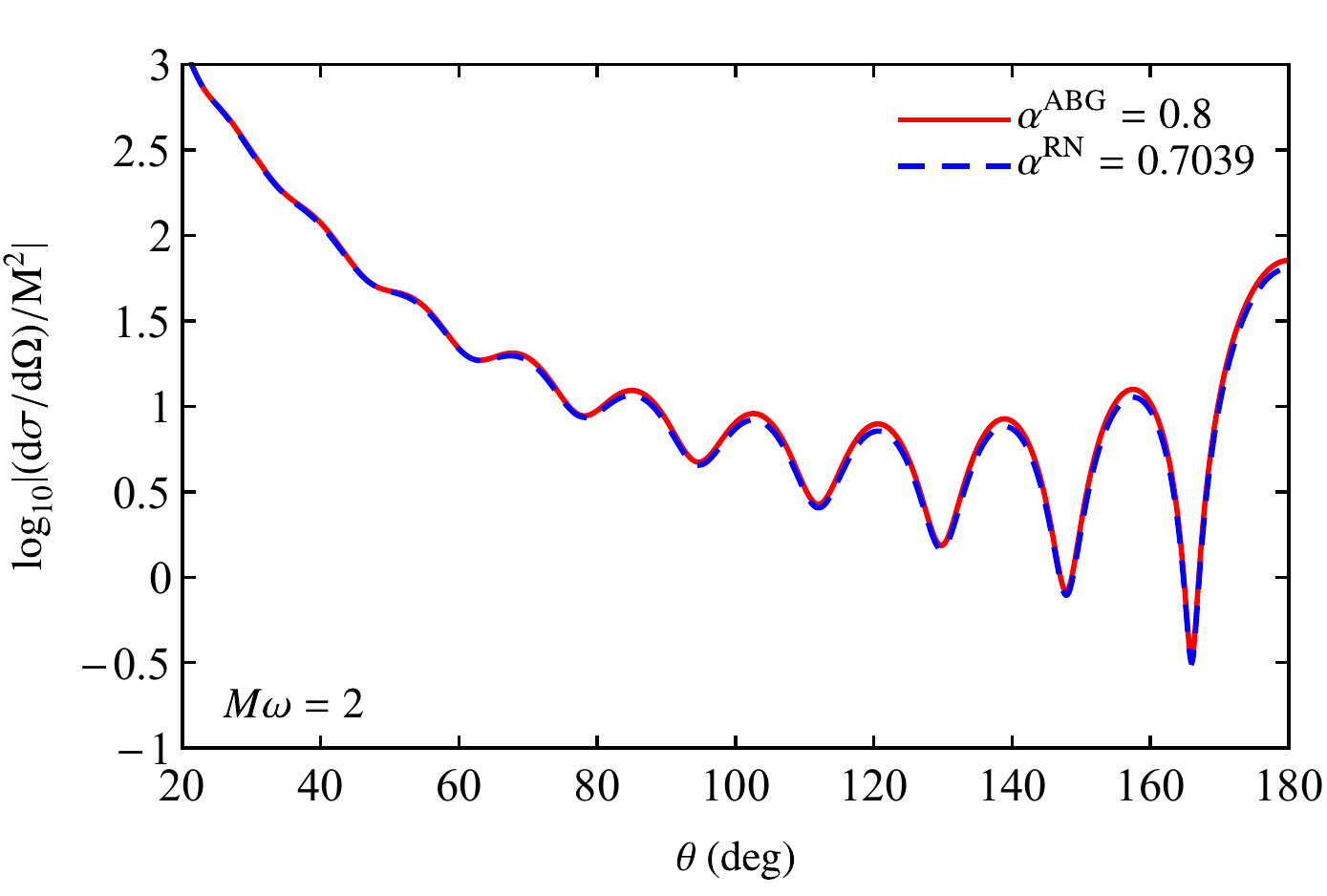}
    \includegraphics[width=\columnwidth]{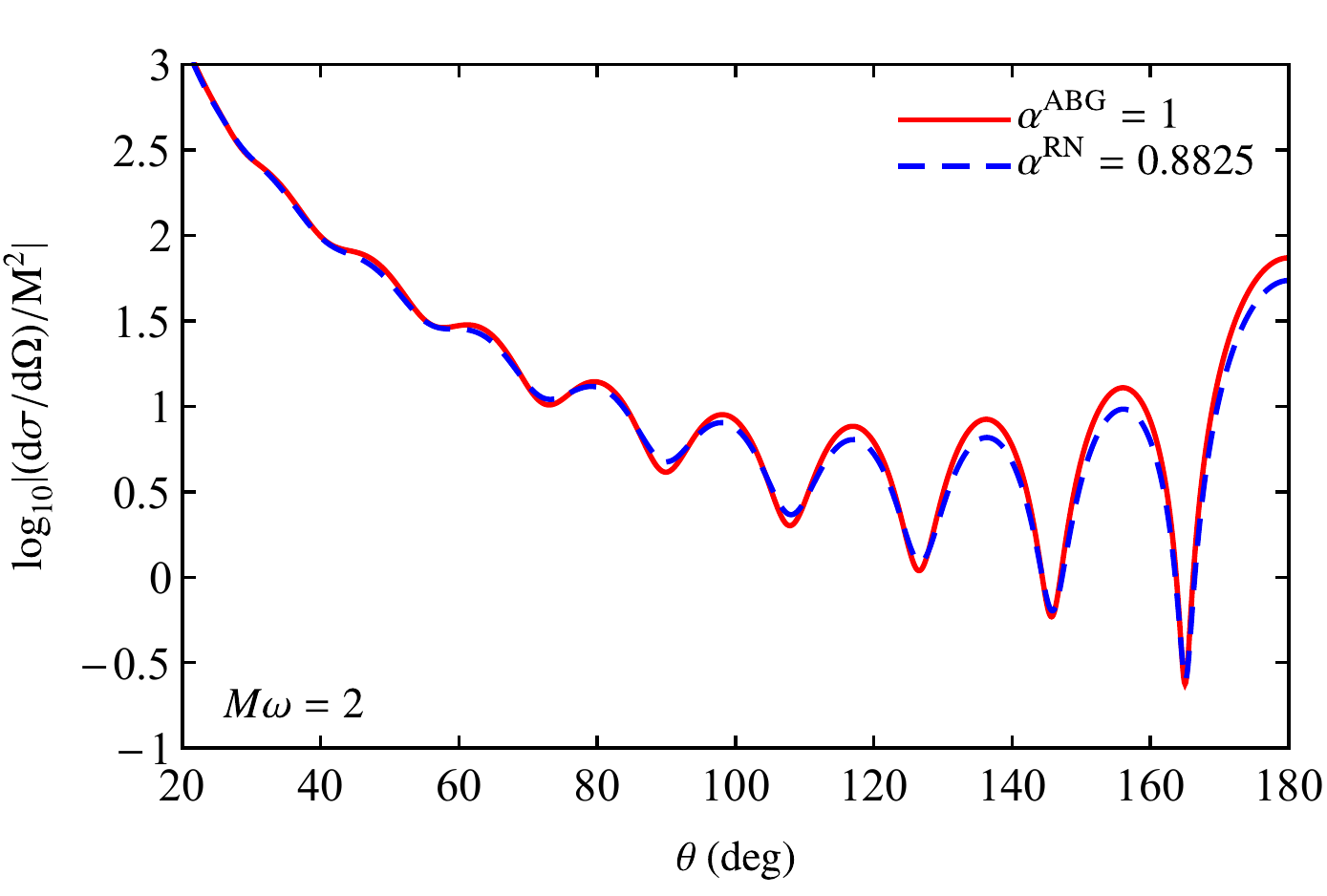}
    \caption{Differential SCS for some pairs ($\alpha^{\rm{ABG}},\alpha^{\rm{RN}}$), considering $M\omega = 2$ in all cases}
    \label{SSCS}
\end{centering}
\end{figure*}

In Fig.~\ref{SSCS} we present the SCSs considering values of the normalized charges for which $b_{g}^{\rm{ABG}}=b_{g}^{\rm{RN}}$. For low-to-moderate values of $\alpha$, we observe that the SCS of ABG and RN BHs can be very similar for arbitrary values of the scattering angle. Remarkably, we see that, although the magnitude of the scattering flux is different for moderate-to-high values of $\alpha$, the oscillatory behavior of the SCSs is still very similar.

\begin{figure*}[!htbp]
\begin{centering}
    \includegraphics[width=\columnwidth]{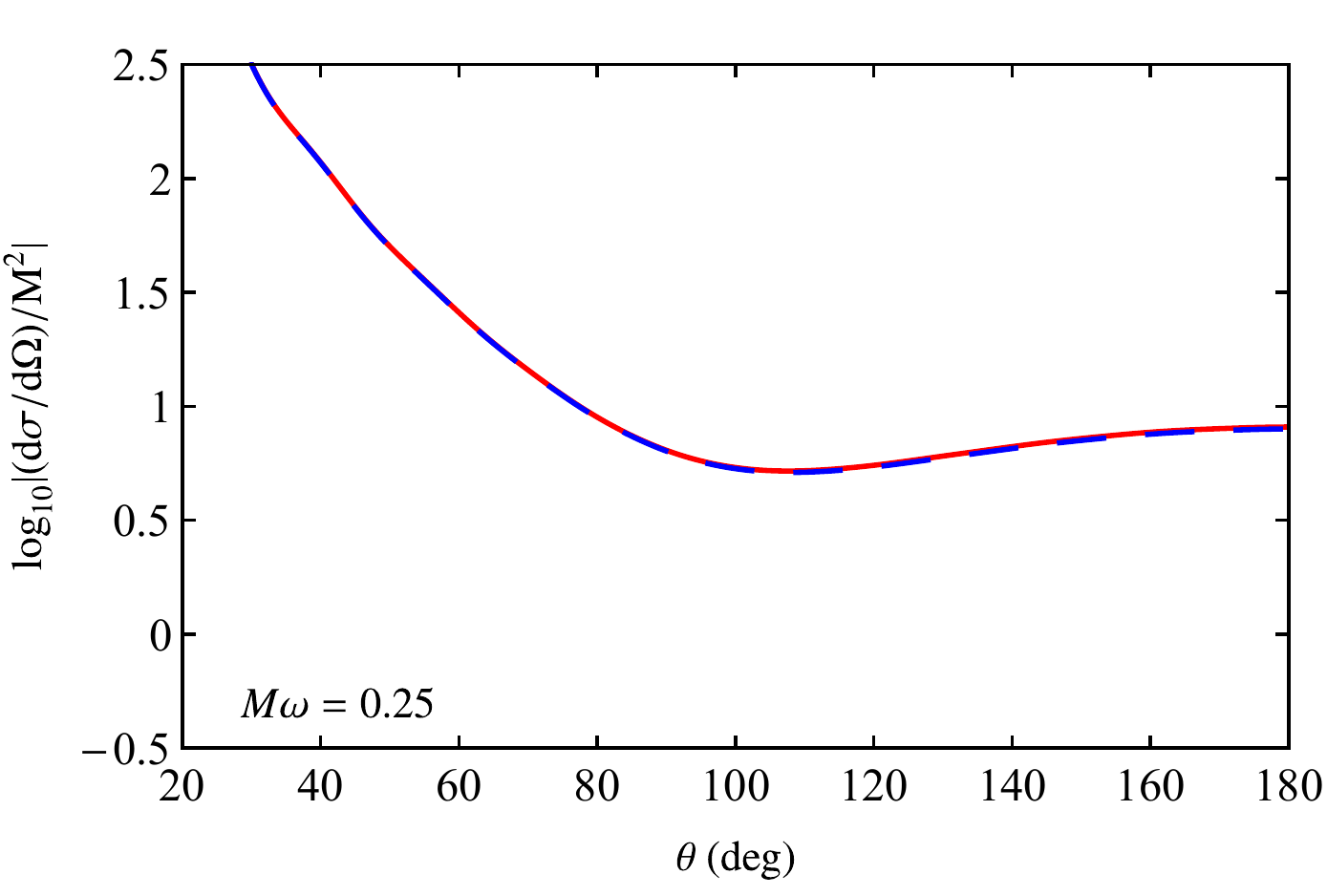}
    \includegraphics[width=\columnwidth]{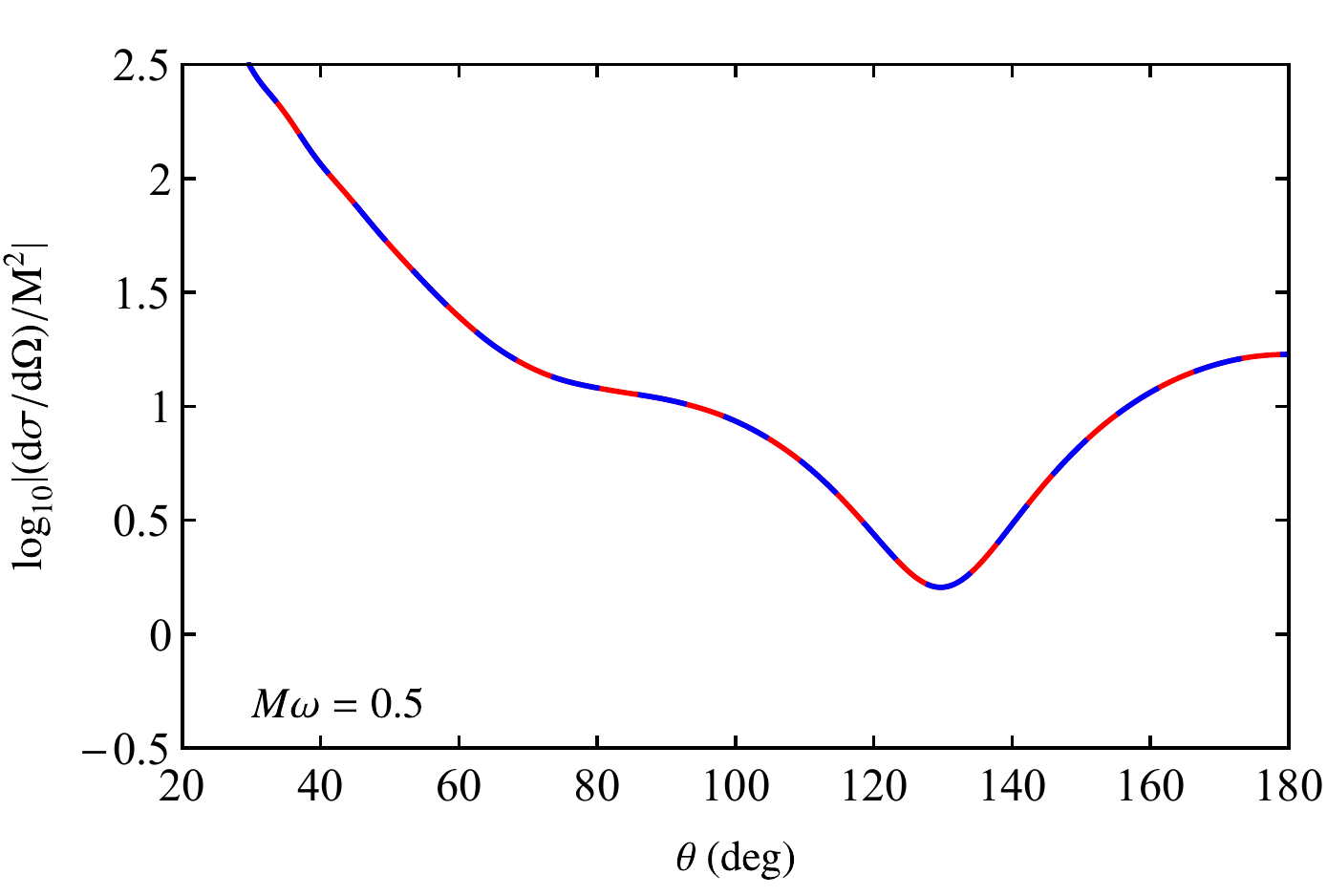}
    \includegraphics[width=\columnwidth]{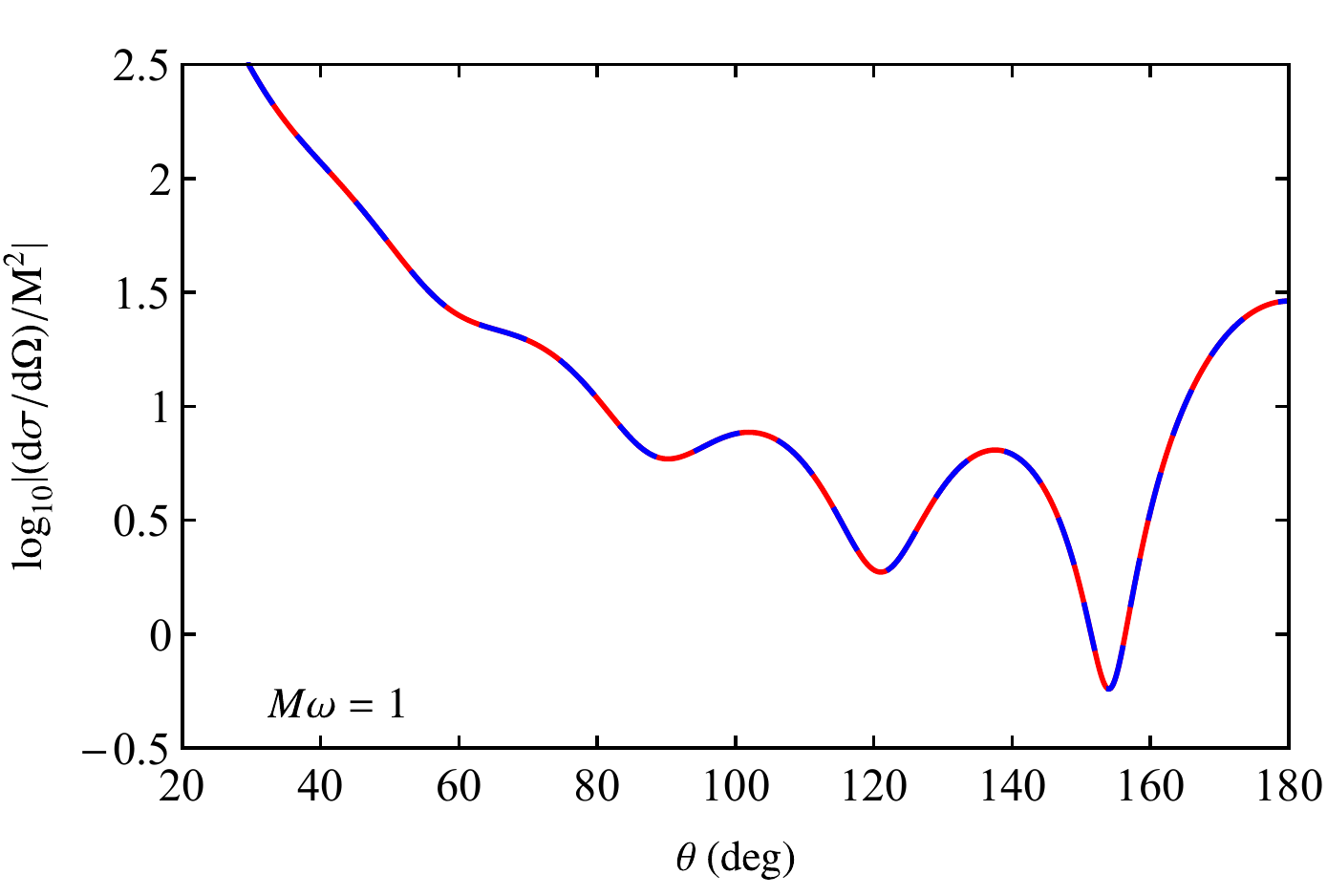}
    \includegraphics[width=\columnwidth]{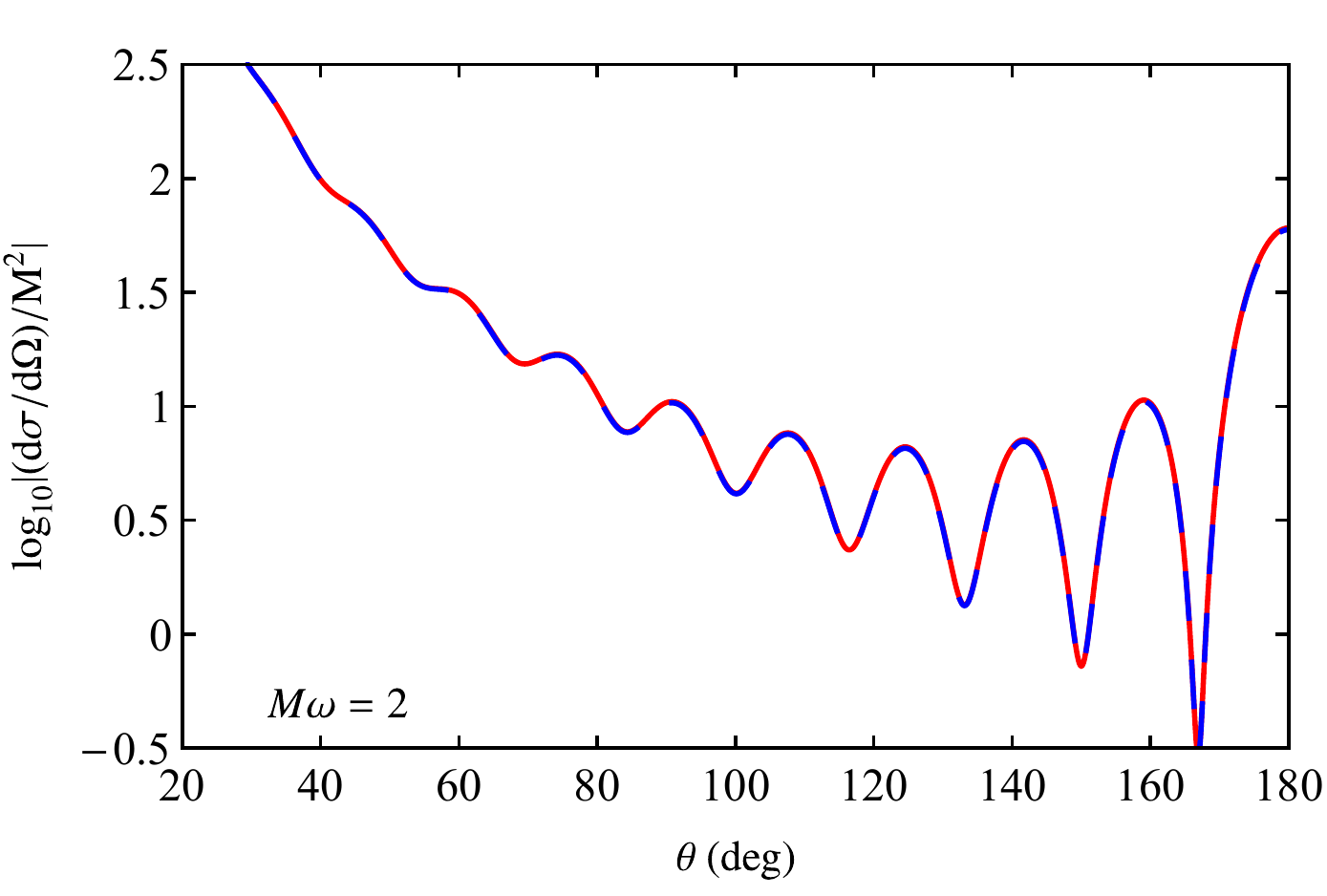}
    \caption{Differential SCS for the pair $\alpha^{\rm{ABG}} = 0.4$ (solid red line) and $\alpha^{\rm{RN}} = 0.3509$ (dashed blue line), which satisfies $b_{g}^{\rm{ABG}} = b_{g}^{\rm{RN}}$, considering distinct values of $M\omega$}
    \label{SSCS2}
\end{centering}
\end{figure*}

In Fig.~\ref{SSCS2} we show the SCS for the pair $\alpha^{\rm{ABG}} = 0.4$ and $\alpha^{\rm{RN}} = 0.3509$, considering different values of $\omega M$. Noticeable, even for small values of $\omega M$, the corresponding SCSs remain very similar. 

\section{Final Remarks}\label{sec:fr}
We have investigated the scattering properties of a massless and chargeless test scalar field in the background of ABG RBHs. We computed the SCS numerically and compared our results with classical (geodesic) scattering and semiclassical glory approximation, showing that: (i) for small scattering angles, the behavior of the SCS is well described by classical results; (ii) for large scattering angles, the SCS can be approximated by the glory formula.

Regarding the influence of electric charge, we have shown that its contribution to the SCS of ABG BHs is more important as we consider higher values of the scattering angle. For small angles, the SCS diverges due to the behavior of the gravitational potential and, in this limit, the contribution of the BH's charge to the SCS is less significant. Moreover, near the backward direction, the interference fringe widths become larger, as one enhances the value of $\alpha$. The latter can be understood with the help of semiclassical formula~\eqref{glory}, which relates the interference fringe widths to $b_g^{-1}$, and we have shown that $b_g$ decreases monotonically with the electric charge for ABG BHs. 

We have found that the equality between the glory parameter $b_{g}$ of ABG and RN BHs can be achieved for $\alpha^{\rm{RN}} \lesssim 0.8825$ and $\alpha^{\rm{ABG}}\leq1$. By considering pairs  $(\alpha^{\rm{ABG}},\alpha^{\rm{RN}})$ for which $b_{g}$ coincide, we have shown that for small-to-moderate values of $\alpha$, the results for the SCSs of ABG and RN BHs are very similar in the whole scattering angle range. Therefore, the results presented along this paper, combined with those obtained in Ref.~\cite{PLC2020}, indicate that, from the point of view of the scattering and absorption properties of massless scalar waves, we may not be able to distinguish RBHs solutions from standard ones.

\begin{acknowledgements}

We are grateful to Funda\c{c}\~ao Amaz\^onia de Amparo a Estudos e Pesquisas (FAPESPA), Conselho Nacional de Desenvolvimento Cient\'ifico e Tecnol\'ogico (CNPq) and Coordena\c{c}\~ao de Aperfei\c{c}oamento de Pessoal de N\'ivel Superior (CAPES) -- Finance Code 001, from Brazil, for partial financial support. This research has also received funding from the European Union's Horizon 2020 research and innovation programme under the H2020-MSCA-RISE-2017 Grant No. FunFiCO-777740. LCSL would like to acknowledge IFPA -- Campus Altamira for the support.

\end{acknowledgements}

\section*{Data availability statement}
	
This manuscript has no associated data or the data will not be deposited. [Authors' comment: There are no associated data available.]

\end{document}